\documentclass{emulateapj}
\usepackage{apjfonts}

\newcommand{\mscl}{$\rm MS\,1054-03$}
\newcommand{\etal}{et al.~}

\slugcomment{Accepted for publication in the Astronomical Journal}

\shorttitle{FIRES: the $\rm MS\,1054-03$ field data}
\shortauthors{F\"orster Schreiber \etal}

\begin{document}

\title{Faint InfraRed Extragalactic Survey: \\
     Data and Source Catalogue of the $\rm MS\,1054-03$ field\altaffilmark{1}}

\author{N. M. F\"orster Schreiber\altaffilmark{2,3},
        M. Franx\altaffilmark{2},
	I. Labb\'e\altaffilmark{4},
        G. Rudnick\altaffilmark{5},
	P.G. van Dokkum\altaffilmark{6},
	G. D. Illingworth\altaffilmark{7},
	K. Kuijken\altaffilmark{2},
	A. F. M. Moorwood\altaffilmark{8},
        H.-W. Rix\altaffilmark{9},
	H. R\"ottgering\altaffilmark{2},
	P. van der Werf\altaffilmark{2}}

\altaffiltext{1}{Based on observations collected at the European
   Southern Observatory, Paranal, Chile (ESO LP Programme 164.O-0612).
   Based on observations with the NASA/ESA Hubble Space Telescope
   obtained at the Space Telescope Science Institute, which is operated
   by the AURA, Inc., under NASA contract NAS5-26555.}
\altaffiltext{2}{Leiden Observatory, PO Box 9513, 2300 RA Leiden,
                 The Netherlands}
\altaffiltext{3}{Present address:
                 Max-Planck-Institut f\"ur extraterrestrische Physik,
                 Giessenbachstrasse, D-85748 Garching, Germany}
\altaffiltext{4}{Carnegie fellow, OCIW,
                 813 Santa Barbara Street, Pasadena, CA 91101}
\altaffiltext{5}{Leo Goldberg Fellow, NOAO,
                 950 N. Cherry Ave., Tucson, AZ 85719}
\altaffiltext{6}{Department of Astronomy, Yale University, P.O. Box 208101,
                 New Haven, CT 06520-8101}
\altaffiltext{7}{UCO/Lick Observatory, University of California,
                 Santa Cruz, CA 95064}
\altaffiltext{8}{European Southern Observatory, Karl-Schwarzschildstr. 2,
                 D-85748 Garching, Germany}
\altaffiltext{9}{Max-Planck-Institut f\"ur Astronomie, K\"onigstuhl 17,
                 D-69117 Heidelberg, Germany}

\begin{abstract}

We present deep near-infrared $J_{\rm s}$, $H$, and $K_{\rm s}$ band imaging
of a field around \mscl, a massive cluster at $z = 0.83$.  The observations
were carried out with the ISAAC instrument at the ESO Very Large Telescope
(VLT) as part of the Faint InfraRed Extragalactic Survey (FIRES).
The total integration time amounts to 25.9~h in $J_{\rm s}$, 24.4~h in $H$,
and 26.5~h in $K_{\rm s}$ band, divided nearly equally between four pointings
covering $5\farcm 5 \times 5\farcm 3$.  The $3\sigma$ total limiting AB
magnitudes for point sources from the shallowest to the deepest pointing are
$J^{\rm tot}_{\rm s, AB} = 26.0 - 26.2$,
$H^{\rm tot}_{\rm AB} = 25.5 - 25.8$, and
$K^{\rm tot}_{\rm s, AB} = 25.3 - 25.7~{\rm mag}$.
The effective spatial resolution of the coadded images has
$\rm FWHM = 0\farcs 48$, $0\farcs 46$, and $0\farcs 52$ in $J_{\rm s}$,
$H$, and $K_{\rm s}$, respectively.  We complemented the ISAAC data with
deep optical imaging using existing Hubble Space Telescope WFPC2 mosaics
through the F606W and F814W filters and additional $U$, $B$, and $V$ band
observations we obtained with the VLT FORS1 instrument.
We constructed a $K_{\rm s}$-band limited multicolour source
catalogue down to $K^{\rm tot}_{\rm s, AB} \approx 25~{\rm mag}$
($\approx 5\sigma$ for point sources).  The catalogue contains 1858 objects,
of which 1663 have eight-band photometry.  We describe the observations,
data reduction, source detection, and photometric measurements method.
We also present the number counts, colour distributions, and photometric
redshifts $z_{\rm ph}$ of the catalogue sources.
We find that our $K_{\rm s}$-band counts at the faint end
$22 \la K_{\rm s,AB} \la 25$, with slope $d\log{N}/dm = 0.20$,
lie at the flatter end of published counts in other deep fields and
are consistent with those we derived previously in the Hubble Deep Field
South (HDF-S), the other FIRES field. 
Spectroscopic redshifts $z_{\rm sp}$ are available for $\approx 330$ sources
in the \mscl\ field; comparison between the $z_{\rm ph}$ and $z_{\rm sp}$
show very good agreement, with
$\langle |z_{\rm sp} - z_{\rm ph}| / (1 + z_{\rm sp}) \rangle = 0.078$.
The \mscl\ field observations complement our HDF-S data set with nearly
five times larger area at $\rm \approx 0.7$ brighter magnitudes, providing
more robust statistics for the slightly brighter source populations.

\end{abstract}


\section{INTRODUCTION}   \label{Intro}

Deep imaging surveys for high redshift studies have, in recent years, been
carried out over an increasing range of wavelengths from the X-ray to the
radio regimes.  An important motivation is the recognition that surveys at
any given wavelength range may provide an incomplete and biased census of high
redshift galaxy populations.  Moreover, a multi-wavelength approach is highly
desirable to assess the nature of and establish links between high redshift
samples selected at various wavelengths, and to constrain the properties of
their stars and gas contents as indicators of their evolutionary state.
Deep surveys at near-infrared wavelengths (NIR; $\rm \lambda = 1 - 2.5~\mu m$)
have emerged as an important way to investigate the high redshift universe
(see, e.g., \citealt{McC04} for a review, and references therein).
NIR observations allow one to access the rest-frame optical light of
galaxies at $z \sim 1 - 4$.  NIR surveys provide a vital complement,
for instance, to optical surveys, which probe the rest-frame UV light
at these redshifts \citep[e.g.][]{Sha01, Pap01}: the rest-frame optical
is less affected by the contribution from young massive stars and by dust
obscuration, and is expected to be dominated by longer-lived lower-mass stars.
Selection in the NIR, compared to the optical, thus provides a more reliable
means of tracing the bulk of the stellar mass at high redshift, of probing
earlier star formation epochs, and of identifying galaxies with spectral
energy distributions (SEDs) similar to those of galaxies locally.
A significant advantage of NIR surveys is that they enable, in the same
rest-frame optical, direct and consistent comparisons between $z \sim 1 - 4$
samples and the well-studied populations at $z \la 1$.

With the sensitive instrumentation and large format NIR detectors mounted
on ground-based $\rm 8 - 10\,m$ telescopes or space-based facilities
available, it is possible to carry out NIR surveys of comparable depth
and quality as in the optical.  This greatly helps in the identification of
counterparts in optical observations and in the construction of accurate
SEDs.  Yet, NIR surveys with sufficient depth and angular resolution
remain challenging.  For instance, a typical normal nearby $L^{\star}$
galaxy would have an apparent $K$-band magnitude
$K_{\rm Vega} \sim 23~{\rm mag}$ if placed at $z \approx 3$ \citep{Kuc01}.
Among current NIR surveys, those reaching such faint limits are still
scarce.  They include the {\em Hubble Space Telescope\/} (HST) NICMOS
fields of the Hubble Deep Field North \citep[HDF-N;][]{Tho99, Dic03} and
Hubble Ultra Deep Field \citep[UDF;][]{Tho05}, the Hubble Deep Field South
(HDF-S) parallel NICMOS field \citep{Wil00}, and the two UDF parallel NICMOS
fields \citep{Bou05}.  Ground-based surveys include the Subaru Deep Field
\citep{Mai01}, and several blank fields imaged at the Keck Telescopes
\citep{Djo95, Mou97, Ber98}.

In this context, we initiated the Faint InfraRed Extragalactic Survey
\citep[FIRES;][]{Fra00} carried out at the ESO {\em Very Large Telescope\/}
(VLT).  The survey consists of very deep imaging in the $J_{\rm s}$,
$H$, and $K_{\rm s}$ bands with the ISAAC instrument \citep{Moo98}
of two fields with existing deep optical HST WFPC2 data: the HDF-S
and \mscl, a field around a massive cluster at $z = 0.83$.
In total, 103~h of ISAAC observations were spent in a single 
$2\farcm 5 \times 2\farcm 5$ pointing for HDF-S and 77~h in a
$5\farcm 5 \times 5\farcm 3$ mosaic of four pointings for \mscl.
The $K_{\rm s}$-band $3\sigma$ limiting total AB magnitudes for point sources
are $\approx 26.2~{\rm mag}$ for HDF-S and $\approx 25.5~{\rm mag}$ for the
\mscl\ field,  and the effective resolution of the coadded maps in $J_{\rm s}$,
$H$, and $K_{\rm s}$ is $0\farcs 45 - 0\farcs 52$ FWHM for both fields.
Together with the HST WFPC2 data and additional VLT FORS1 optical imaging
for the \mscl\ field, this provides a unique data set with seven or eight
band photometry covering $0.3 - 2.2~{\rm \mu m}$ over a total area of
$\approx 35~{\rm arcmin^{2}}$.  With the HDF-S data, FIRES reaches fainter
limits in all three NIR bands than any other existing ground-based survey
and provides the deepest $K$-band map to date.  The \mscl\ field is an
essential complement as it considerably increases the survey area to an
$\approx 0.7~{\rm mag}$ brighter limit.

The data and source catalogues of the HDF-S are described by \citet{Lab03a}.
Several key results of both fields have been published, including SED
modeling, spectroscopy, and clustering analysis of the new population
of ``Distant Red Galaxies'' at $z > 2$
\citep{Fra03, Dok03, Dok04, Dad03, FS04, Lab05},
determination of the stellar rest-frame optical luminosity density,
colours, and mass density at $z = 0 - 3$ \citep{Rud01, Rud03},
determination of the rest-frame optical sizes of galaxies out to $z \sim 3$
\citep{Tru03, Tru05}, and the discovery of six large disk-like galaxies at
$z = 1.4 - 3.0$ \citep{Lab03b}.
In this paper, we present the data and source catalogues of the \mscl\ field.
As for HDF-S, the \mscl\ raw and reduced data sets are available from the
FIRES Web site\footnote{http://www.strw.leidenuniv.nl/\~{ }fires}.
Sections~\ref{Sect-obs} and \ref{Sect-red} describe
the observations and data reduction.
Sect~\ref{Sect-images} discusses the final reduced images.
Section~\ref{Sect-det} describes the $K_{\rm s}$-band limited
source detection and photometric measurements methods and
\S~\ref{Sect-redshifts} the photometric redshift determinations.
Section~\ref{Sect-catalogue} lists the catalogue parameters.
Section~\ref{Sect-analysis} presents the completeness analysis,
number counts, magnitude and colour distributions of the
$K_{\rm s}$-band selected sources.
Section~\ref{Sect-conclu} summarizes the paper.
All magnitudes are expressed in the AB photometric system
\citep{Oke71} unless explicitely stated otherwise.

\section{OBSERVATIONS}    \label{Sect-obs}

\subsection{The $\mathit{MS\,1054-03}$ Field}   \label{Sub-obs_field}

Our strategy for the FIRES survey was to complement our very deep NIR
imaging of HDF-S with a wider field, allowing for better statistics on
source populations down to flux levels twice brighter than the faint limits
of HDF-S.  The \mscl\ field was ideally suited for this purpose because of
available HST WFPC2 imaging offering a unique combination of field size
and depth.  The WFPC2 data consist of a mosaic of six pointings covering
$\approx 5^{\prime} \times 5^{\prime}$ observed through the F606W and F814W
filters (effective wavelengths $\lambda_{\rm eff} = 6030$ and 8040~\AA, and
effective widths $\Delta\lambda = 1500$ and 1540~\AA, respectively).
\footnote{All bandpass wavelengths and widths quoted in this paper refer
to the effective transmission profile accounting for the filters and all
other instrumental optical components, the telescope mirrors, the detectors,
and the Earth's atmosphere.}
We will denote these bandpasses as $V_{606}$ and $I_{814}$.  The total
integration time was 6500~s per band and pointing.  To extend the optical
wavelength coverage, we obtained $U\,B\,V$ imaging with the VLT FORS1
instrument.

The \mscl\ field benefits from an extensive amount of existing data over
the electromagnetic spectrum, including a deep mosaic at 850~\micron\ taken
with the Submillimeter Common-User Bolometric Array (SCUBA; K. K. Knudsen
\etal, in preparation), deep radio 5~GHz imaging from the Very Large Array
\citep[VLA;][]{Bes02}, and X-ray observations with the ACIS camera onboard
the Chandra Observatory \citep{Jel01, Rub04}.
Optical imaging with ACS onboard HST has also now become publicly available
\footnote{The ACS data have been obtained too recently to be included in
the work presented here, which has extended over several years and focusses
on the initial phase of FIRES.}.
Spectroscopic redshifts have been measured for over 400 galaxies in the
field from data collected at the Keck and Gemini telescopes and at the VLT
(\citealt{Tra99, Tra03, Dok00, Dok03, Dok04}; S. Wuyts \etal, M. Kriek \etal,
in preparation).  These spectroscopic determinations enabled us to calibrate
our photometric redshifts (\S~\ref{Sect-redshifts}).

The $z = 0.83$ \mscl\ cluster itself has but a small influence over the area
surveyed.  The majority of galaxies detected in the mosaics lie in front of or
behind the cluster.  The cluster mass distribution has been modeled accurately
by \citet*{Hoe00} from weak lensing analysis, so that the gravitational lensing
can be well accounted for.  The average background magnification effects over
the field of view covered by the FIRES observations range from 5\% to 25\%
between $z = 1$ and 4, and are largest in the cluster's central regions.
We emphasize that colours and SEDs are not affected since gravitational
lensing is achromatic.  Moreover, the surface brightness is preserved and,
to first order, the surface density to a given apparent magnitude is
insensitive to lensing (the exact effect depends on the slope of the
luminosity function of the sources under consideration).

Table~\ref{tab-obs} summarizes the properties of the \mscl\ data set and
Figure~\ref{fig-fovs} illustrates the fields of view of the different
instruments.  The observations and reduction of the HST WFPC2 mosaics are
presented by \citet{Dok00}.  In this and the following section, we focus on
the data taken with ISAAC and FORS1, both mounted at the VLT Antu telescope.
All VLT observations were acquired in service mode as part of the ESO Large
Program 164.O-0612 (P.I. M. Franx).

\subsection{ISAAC Near-infrared Observations}   \label{Sub-obs_isaac}

The \mscl\ field was observed with ISAAC \citep{Moo98} in imaging
mode through the $J_{\rm s}$, $H$, and $K_{\rm s}$ filters
($\lambda_{\rm eff} = 1.25$, 1.65, and 2.16~\micron, and
$\Delta\lambda = 0.29$, 0.30, and 0.27~\micron, respectively).
The observations were carried out on 36 nights, in 1999 December, 2000 May,
and 2001 February through June.  The short-wavelength camera of ISAAC is
equipped with a Rockwell Hawaii $1024 \times 1024$ HgCdTe detector array and
provides a scale of $\rm 0\farcs 148~pixel^{-1}$ for a total field of view of
$2\farcm 5 \times 2\farcm 5$.  The WFPC2 mosaic was covered with a square of
$2 \times 2$ pointings (referred to as F1 to F4; see Figure~\ref{fig-fovs}).
Nominal pointings ensured that adjacent fields overlap by
$\approx 10^{\prime\prime}$.

The data were obtained in ``observing blocks'' (OBs) consisting of sequences
of 30 exposures totaling 1~h integration time in $J_{\rm s}$ and $H$, and 40
exposures totaling 40~m in $K_{\rm s}$.  Successive exposures within each OB
were dithered randomly in a 20\arcsec\ box to allow the construction of sky
frames with minimal object contamination.  Typical exposure times in
$J_{\rm s}$, $H$, and $K_{\rm s}$ were of 120, 120, and 60~s, split into 4,
6, and 6 sub-integrations, respectively.  The entire data set includes 3207
frames grouped in 93 OBs, with total integration times of $\rm 6-7~h$ per
field and per band.  We rejected about 3\% of the frames, representing 3\%
of the total integration time, because they did not meet our atmospheric
seeing or transmission requirements, or because they were affected by strong
instrumental artefacts.  Throughout the paper, we consider only the 3104
frames from 92 OBs with sufficient quality.

Most of the data were obtained under excellent seeing, with median
full-width at half maximum $\rm FWHM = 0\farcs 47$ in $J_{\rm s}$ and
$0\farcs 45$ in both $H$ and $K_{\rm s}$.  Ninety percent of the data have
a seeing better than $0\farcs 65$.  Figure~\ref{fig-seeing} gives the seeing
distribution of the individual frames.  The FWHM was determined by fitting
Moffat profiles to bright, isolated, and unsaturated stars in each image
(mostly three to five depending on the field, on the signal-to-noise ratio
of the stars, and on whether they lie within the field of view of individual
dithered frames), and averaging the FWHMs per image.  The conditions were
photometric for about half of the observing nights, or 53\% of the OBs.

In the course of the 19-month interval during which the NIR observations
were taken, the Antu primary mirror was twice re-aluminized, in 2000 February
and 2001 March.  We will refer to the three periods separated by these events
as periods P1, P2, and P3.  The effects of re-aluminization are noticeable in
our data, with an increase in sensitivity between successive periods as deduced
from bright stars counts in the \mscl\ data and reflected in the photometric
zero points.  Based on the standard stars data used for flux calibration
(\S~\ref{Sub-red_isaac}), the increase in $J_{\rm s}$, $H$, and $K_{\rm s}$
amounts to 37\%, 47\%, and 50\% between P1 and P2, and 29\%, 26\%, and 19\%
between P2 and P3, for a net gain of $\approx 80\%$.  Seven OBs were taken
in P1, 41 in P2, and 45 in P3.  The ISAAC pixel scale also varied slightly
over time, by a maximum of 0.27\%.

Figure~\ref{fig-obs_param} shows the variations with time and airmass of the
relative count rates of stars and of the background levels in the individual
frames.  The stars are the same as used for the seeing measurements.
Their counts were integrated in a 6\arcsec -diameter circular aperture in
the sky-subtracted frames and normalized to the median values in period P3.
The background was computed as the median count rates over the raw frames.
The figure also plots the seeing FWHM versus airmass.

The stars count rates increase sharply after each realuminization and,
over period P2, are consistent with the long-term progressive loss of
sensitivity between two realuminizations.  In contrast, the background
levels have overall remained fairly constant, with no systematic trend
with period and the variations reflecting mostly the nightly observing
conditions.  This constancy in average background levels was also noticed
in the HDF-S data set and presumably indicates that light is scattered
in as much as out of the beam.  It implies a substantial increase in
telescope efficiency and in source signal-to-noise ratio (S/N) achievable
in background-limited data after realuminization.  

The NIR background can vary on timescales of minutes; it is dominated by
telluric OH airglow lines in the $J_{\rm s}$ and $H$ bands, and by thermal
emission from the telescope, instrument, and atmosphere in the $K_{\rm s}$
band.  The nightly variations in our data are generally largest and most
rapid in the $H$ band and show little correlation with the hour of observation.
The $K$-band background has comparable levels but smaller short-term variations
and is systematically higher at the beginning and end of the night.  The
$J_{\rm s}$ background is the lowest and most stable, and peaks at the
beginning of the night.

Variations in stars count rates within each period reflect changes in sky
transparency and there is no apparent airmass dependence.  This indicates
that the atmospheric extinction is negligible for airmasses of $1.05 - 2.2$
as covered by the observations.  The seeing is expected to correlate with
airmass (roughly as $\rm FWHM \propto [airmass]^{0.6}$)
but no obvious trend is seen in our data, partly due to the small range of
airmass and because the seeing is largely determined by the actual observing
conditions.  The data suggest an overall increase in background levels with
airmass, most clearly in $H$ and $K_{\rm s}$, which could be explained by the
increasing path length of the thermal and OH-line emitting layers.  However,
for successive observations such as within OBs, variations in airmass are
directly coupled with time so that part of possible trends with airmass
may be attributed to nightly patterns in the background.

\subsection{FORS1 Optical Observations}   \label{Sub-obs_fors}

The \mscl\ field was observed with FORS1 \citep{App98} in direct imaging
mode on six nights between 2000 January and March using the Bessel $U$,
$B$, and $V$ filters ($\lambda_{\rm eff} = 3680$, 4330, and 5550~\AA, and
$\Delta\lambda = 310$, 870, and 1070~\AA, respectively).  FORS1 employs a
$\rm 2048 \times 2048~pixels$ Tektronix CCD detector.  The standard
resolution collimator was selected for the observations, giving a 
$\rm 0\farcs 20~pixel^{-1}$ scale and a $\rm 6\farcm 8 \times 6\farcm 8$
field of view covering entirely the ISAAC and WFPC2 mosaics.
The data were collected in OBs of six successive single exposures of 550, 400,
and 300~s in $U$, $B$, and $V$.  The total integration times are 4.6, 2.0, and
0.5~h, respectively, distributed over 5, 3, and 1 OBs.  This excludes the few
aborted or complete OBs with insufficient quality, lying well outside our
observing conditions specifications.  Within each OB, the exposures were
dithered in a box of $\approx 40^{\prime\prime} \times 30^{\prime\prime}$.

The average seeing FWHM of the data is $0\farcs 64$ in $U$, $0\farcs 55$ in
$B$, and $0\farcs 67$ in $V$.  The seeing was determined as for the ISAAC data
by fitting Moffat profiles to five bright, isolated, unsaturated stars in the
individual frames.  The $B$- and $V$-band data were all taken prior to the 2000
February realuminization while the $U$-band observations were carried out both
before and after.  Our \mscl\ data indicate an increase in $U$-band sensitivity
of $\approx 20\%$ based on stars count rates in a 6\arcsec -diameter aperture
and small rotation angle differences in the camera orientation
($0.15^{\circ}-0.57^{\circ}$ between various OBs).  The data were obtained
during photometric nights except for the three $U$-band OBs taken in 2000 March.

\section{DATA REDUCTION}   \label{Sect-red}

\subsection{ISAAC Data Reduction}   \label{Sub-red_isaac}

We reduced the ISAAC data using the 
DIMSUM\footnote{DIMSUM is the Deep Infrared Mosaicking Software
package developed by Peter Eisenhardt, Mark Dickinson, Adam Stanford,
and John Ward, and is available via ftp to
ftp://iraf.noao.edu/iraf/contrib/dimsumV2/dimsum.tar.Z}
package within
IRAF\footnote{IRAF is distributed by the National Optical Astronomical
Observatories, which are operated by AURA, Inc., under cooperative
agreement with the National Science Foundation.},
the ECLIPSE\footnote{ECLIPSE is a software package written by N. Devillard,
which is available at http://www.eso.org/projects/aot/eclipse.}
software, and additional routines we have developed specifically for
our FIRES observations.  We followed the general procedure for the HDF-S
described by \citet{Lab03a} with adjustments necessary because of the
different characteristics of the \mscl\ data set.  In particular, the
observations were taken over a longer period where the instrument behaviour
changed noticeably.  Also, the shallower data and more crowded field around
\mscl\ required optimization of various steps to minimize residual noise
and systematic effects.

\subsubsection{Dark, Bias, and Flat Field}

We generated dark$+$bias images for use in making flat fields by averaging
together individual dark frames taken with each detector integration time
(DIT) employed for the flat fields.  The typical level of the dark frames
for a given DIT varied significantly over time with five distinct epochs
identified in our data.  We thus created dark$+$bias images appropriate
for each epoch in addition to those made from all available dark frames.

We constructed flat fields from sequences of successive exposures of the sky
taken during twilight, on nights where the \mscl\ field was observed or as
close as possible in time, using the ECLIPSE task ``flat.''  This task also
produced the initial bad (dead) pixel masks.  We made three flat fields per
night and band as follows:
(1) by forcing ECLIPSE/flat to compute the dark$+$bias by itself,
(2) by providing the average dark$+$bias image from all available
data for the corresponding DIT, and 
(3) by providing the dark$+$bias image for the corresponding DIT
and observing epoch.

The flat fields often exhibited abrupt jumps at mid-array (row 512) with
amplitude up to 20\% and significant differences in large-scale gradients
between the three methods, typically at the 5\% level.  These features carry
the imprint of the detector bias, which dominates the signal in the dark
frames.  The bias level is very uniform along rows but varies importantly
along columns, with a steep initial decline flattening progressively from
bottom to top of each of the lower and upper halves of the array frame.
The bias depends not only on the DIT but also on the detector illumination
during and prior to an exposure, so that for a given DIT the dark$+$bias
image may not represent well the dark$+$bias present in the twilight frames.
For each night and band, we chose the flat field for which the bias signatures
were minimized.  This was generally the one obtained with method 1, which
better accounted for the intrinsic dark$+$bias of the twilight frames, but
in a few cases method 3 was the most successful.  The nightly flat fields
also varied noticeably between epochs where the dark$+$ bias changed most
but were otherwise very similar.  We created ``master flat fields'' by
averaging the adopted nightly flat fields for each epoch.  All flat fields
were normalized to unit mean over the image.

\subsubsection{Sky Subtraction, Cosmic Rays Rejection, and OB Combination}

For each science exposure in a given OB, we constructed a ``sky'' image
to remove the sky and telescope background, and the dark$+$bias signal.
We used a minimum of three and a maximum of eight temporally adjacent frames
depending on when the science exposure was taken during the OB sequence.
In a ``first pass,'' the individual sky frames were scaled to a common
median level equal to that of the science frame, averaged together with
minimum-maximum rejection algorithm at each pixel, and subtracted from
the science frame.  Cosmic rays and hot pixels were identified using the
``L.~A. Cosmic'' routine \citep{Dok01} and negative outliers at $> 8\sigma$
below the mean residual background were added to the bad pixel lists.
The sky-subtracted frames within each OB were aligned using integer-pixel
shifts and coaveraged.  From this intermediate combined OB image, an object
mask was created by applying an appropriate threshold to identify astronomical
sources.  In a second ``mask pass,'' the sky subtraction procedure was
repeated using the object masks to minimize the contribution of sources in
the background computation.  The mask-pass sky-subtracted frames were divided
by the master flat field for the corresponding epoch of observation and band,
and combined as before.  Associated weight maps were generated, proportional
to the total integration time of all frames contributing at each pixel.
We normalized the final OB images to a 1\,s integration time and the
weight maps to a maximum value of~1.

Non-linearity (raw counts exceeding 10000~ADU, for which ESO reports
99\% linearity) is not a concern except for the very few brightest sources
($\leq 3$ per ISAAC field) in the frames obtained under the best seeing and
transparency conditions.  We did not apply any correction for non-linearity
but excluded those stars with raw counts well above 10000~ADU in the flux
calibration and PSF determination described below.

In coaveraging the reduced frames, bad pixels and cosmic rays were masked
out and no clipping or rejection algorithm was applied.  We verified that no
strongly deviant pixel, either persistent or as single event, was missed by
comparing the OB images to versions obtained by median-combining the registered
sky-subtracted frames without masking and from the behaviour of individual
pixels throughout the stack of unaligned unmasked sky-subtracted frames.
In several OBs, we noticed satellite tracks, moving targets (presumably
Kuiper Belt objects), and spurious features produced by reflection effects
of bright sources falling on the edges of the detector array.  We removed
them with customized masking.  We inspected the individual reduced frames to
identify those with significant residual artefacts due to poor sky subtraction
or instrumental features.  We implemented additional mask-pass routines in
DIMSUM to improve the quality of the affected data, described in what follows.

{\em Bias residuals ---\/}
Particular processing was required to account for the time and
illumination-dependent bias level of the detector.  The median-scaling
applied in constructing the sky images accounts for the global intensity
of the dark$+$bias but not for its spatial structure.  Because the bias
variations are strong along columns but negligible along rows, we removed the
residuals by subtracting the median row by row in all sky-subtracted frames.
In many OBs, the bias varied so strongly and rapidly that it led to spurious
``numerical ghosts'' with shapes reminiscent of the masked sources.  These
features were most conspicuous in regions of steepest spatial gradients in
the bias (lower parts of the bottom and top half of the frames) and during
periods of steepest gradients in time (related to the illumination history).
The average of the scaled sky frames did not represent accurately the local
background in the science frame and the effect was worsened by the object
masking which reduced locally the number of frames used to estimate the
background.  The computed background in the regions with objects masked
in one or more of the sky frames thus differed significantly from that
of adjacent regions masked in none of the sky frames, imprinting
positive ghosts for a locally underestimated background or vice-versa.
We eliminated the numerical ghosts while constructing the sky frames:
in addition to the median-scaling, we accounted for an additive component
representing the bias offsets between a science frame and each of its
corresponding scaled sky frames by fitting a median to each row in the
difference maps.  We applied this procedure to the 28 OBs where ghosts
were clearly recognized by eye inspection, mostly in the $J_{\rm s}$
and $H$ bands.

{\em Residual large-scale gradients ---\/}
For about half of the data, significant large-scale gradients remained in
the sky-subtracted frames.  They were caused primarily by highly variable
sky and thermal background emission, and stray light.
The fractions of OBs affected in each band reflect the background
behaviour seen in Figure~\ref{fig-obs_param}: 92\% in $H$, 46\% in $K_{\rm s}$,
and 27\% in $J_{\rm s}$.  We first attempted to remove these residuals by
fitting two-piece cubic splines directly to the object-masked frames, along
rows or columns depending on the gradient pattern.  Owing to crowding in the
\mscl\ field, the fits were however generally poorly constrained.
We thus created template ``flat images'' by registering and coaveraging all
sky-subtracted frames unaffected by residual gradients, taken from all OBs
of the appropriate field and band to increase the S/N ratio.  We subtracted
the flat templates from the affected frames.  This largely eliminated sources,
leaving residual cores due to small mismatches in alignment and seeing.
We used these difference maps to create new masks and fit the two-piece
cubic splines.  Because the masked area was substantially reduced, the fits
were much better constrained, allowing successful removal of the residual
gradients.  We performed this second-order sky subtraction for the 29 OBs
where the gradients were strongest and apparent in the combined OB images.

{\em Instrumental features and other artifacts ---\/}
A high spatial frequency pattern known as the ``odd-even column effect''
appeared in part of the data taken after the 2001 March realuminization
(19 OBs).  It is characterized by an offset in level between adjacent columns,
occasionally further modulated over a width of about 20 pixels and generally
different between detector quadrants.  When present, we corrected for this
artifact by subtracting the median of the background along columns and for
each quadrant separately in the object-masked sky-subtracted frames.

Finally, high to moderate spatial frequency ripples presumably from
moonlight reflections were present in the background of 41 OBs, with
important frame-to-frame variations in intensity, structure, and orientation.
A semi-circular high-frequency time-dependent interference pattern was also
obvious in 3 OBs.  The complexity and variability of these artifacts make
them very difficult to correct for.  Since they generally averaged out in
the combined OB images, we did not treat them in any particular way except
by excluding the most affect frames which left noticeable patterns in the
OB images.

\subsubsection{Rectification}

We corrected the OB images for the documented ISAAC geometrical distortion
while simultaneously registering and interpolating the ISAAC data to the
WFPC2 maps at $\rm 0\farcs 1~pixel^{-1}$.  This ``rectification'' was done
using the same algorithm as for the HDF-S data described by \citet{Lab03a}.
Briefly, the transformation corrects for distortion, aligns the ISAAC OB
images (allowing for sub-pixel shifts), and applies the best-fit linear
transformation determined from the positions of stars and point-like sources
across the $J_{\rm s}$-band images and the WFPC2 $I_{814}$-band mosaic.
The images are resampled using third-order
polynomial interpolation, and only once to minimize the effects on the noise
properties.  We rescaled the rectified OB images to ensure flux conservation.
Since the ISAAC field distortions have small amplitude and thus negligible
effects on the photometry, we used a constant flux scaling factor for the
entire images.

The rectification of the OB images instead of the individual
frames affects the point-spread function (PSF) because of the dithering
and integer-pixel registration within each OB.  Figure~\ref{fig-seeing_OBs}
plots the PSF FWHM measured in the rectified OB images versus the median of
the FWHMs in the single frames of the corresponding OB.  The effect is
negligible at $\rm FWHM \ge 0\farcs 5$ and increases towards lower FWHM
up to a maximum of 25\% at $0\farcs 3$, the limit of undersampling.

\subsubsection{Flux Calibration}

We determined the instrumental photometric zero points (ZPs) using
observations of standard stars.  All stars were selected from the
LCO/Palomar NICMOS list \citep{Per98} except for one taken from the
UKIRT Faint Standards list \citep[][observed on 1999 December 18]{Cas92}.
We measured all fluxes and magnitudes for photometric calibration in a
6\arcsec -diameter circular aperture, which proved to be adequate by
monitoring the growth curve of the standard stars and of reference stars
in the \mscl\ field.

The standard stars were observed in sequences of five dithered exposures.
We integrated the count rates of the stars within the 6\arcsec -diameter
aperture in each frame of a sequence after sky subtraction and flat-fielding
(using the appropriate flat field for the epoch of observation and band).
We computed single-star fluxes by averaging, with $2\sigma$-clipping, the
count rates for each sequence.  Typical uncertainties, taken as the median
of the standard deviations of single-star fluxes, were of 1.7\% (average of
2.1\%).  Using the calibrated magnitudes, we derived single-star ZPs.
We identified as non-photometric the nights where single-star ZPs varied
by more than twice the typical uncertainties (i.e. 3.4\%), and excluded
them from the subsequent analysis.  For each of the remaining nights, we
derived nightly ZPs by averaging the single-star ZPs.  Because the nightly
ZPs exhibit a large systematic increase after realuminization events, we
computed period ZPs by averaging with $2\sigma$-clipping the nightly ZPs
for each of P1 to P3.  Nights rejected by the $2\sigma$-clipping were
added to the initial list of non-photometric nights.  In total, 17 out
of 36 nights were photometric according to our criteria.

The median night-to-night dispersion of the ZPs is 0.024~mag and varies from
0.007 and 0.060~mag depending on the band and period.  The average ZPs per
period in $J_{\rm s}$, $H$, and $K_{\rm s}$ increased by 0.340, 0.415, and
0.441~mag from P1 to P2, and by 0.276, 0.255, and 0.188~mag from P2 to P3.
We adopted as reference ZPs for our \mscl\ data those of P3, which are
reported in Table~\ref{tab-zps}.

We did not account for atmospheric extinction or for colour terms.
The calibration stars observations cover airmasses $1 - 1.6$, at which 95\%
of the \mscl\ data were obtained.  As for the \mscl\ stars count rates
(Figure~\ref{fig-obs_param}), the standard stars data do not show any
measurable correlation with airmass.
The ISAAC $H$ and $K_{\rm s}$ filters match well those used to establish the
``LCO'' standard system of \citet{Per98} while the $J_{\rm s}$ filter has a
slightly redder effective wavelength than the LCO $J$ filter.  For a colour
term $\gamma\,(J - K)_{\rm LCO}$, the colour coefficient $\gamma$ is
theoretically expected to be $\rm < 0.01~mag$ in $H$ and $K_{\rm s}$,
and $\rm -0.04~mag$ in $J_{\rm s}$ \citep{Ami02} but these values have
not been verified experimentally yet. 
The ZPs derived for the individual stars do not reveal any dependence on
$J-K$ colour over the range $\rm 0.24 - 0.46~mag$ covered by the standard
stars set, with a scatter much larger than the expected colour effects.

We calibrated the \mscl\ rectified OB images using a set of $4 - 5$ bright,
isolated, and unsaturated stars for each field.  We first considered only the
OBs taken during photometric nights.  We integrated the stars count rates in
a 6\arcsec -diameter aperture and applied the nightly ZPs to obtain individual
magnitudes.  We adopted as calibrated magnitudes the $2\sigma$-clipped averages
of all individual measurements.  We then included all OBs and computed for each
star and OB the appropriate scaling factor based on the star's magnitude and
the adopted ZPs of Table~\ref{tab-zps}.  We calibrated the OB images 
individually using the $2\sigma$-clipped average of the scaling factors
of all stars per OB.

For each reference star in the \mscl\ field, we computed the $1\sigma$
dispersion of their calibrated magnitudes in the photometric OBs.
The median of the dispersions of all reference stars is 0.030, 0.032,
and 0.025~mag in $J_{\rm s}$, $H$, and $K_{\rm s}$, respectively.
We adopt 0.03~mag as the internal accuracy of the photometric
calibration of the NIR \mscl\ data set.

\subsubsection{Final Mosaicking}

We produced the final $J_{\rm s}$-, $H$-, and $K_{\rm s}$-band images of
\mscl\ in two steps: we first combined all OB images per field and then
combined the four field maps into the large mosaic.  We proceeded in this
way because of systematic differences in PSF of the OB images between the
data sets of each field, which would be problematic for the overlap areas.

To create the separate field maps, we weight-averaged the calibrated
rectified OB images per band, without clipping or rejection algorithm.
The weight applied at each pixel $w_{\rm pix}$ was proportional to the
integration time $t_{\rm int,pix}$, scaled by the inverse of the
pixel-to-pixel variance of the background noise in each image
$\overline{\sigma}_{\rm OB}^{~2}$ and of the squared $\rm FWHM_{OB}$
of the OB PSF:
\begin{equation}
w_{\rm pix} \propto 
\frac{t_{\rm int,pix}}
     {\overline{\sigma}_{\rm OB}^{~2}\,{\rm FWHM}_{\rm OB}^{~2}}.
\label{Eq-weight}
\end{equation}
The $t_{\rm int,pix}$ term corresponds to the weight maps of the OBs
multiplied by their respective total integration time, and thus accounts
for pixels that were excluded because bad/hot or affected by other artefacts.
The background noise was estimated on source-free areas of the calibrated
but pre-rectified OB images to avoid the spatially-correlated noise
introduced by rectification.
The $\overline{\sigma}_{\rm OB}^{~2}\,{\rm FWHM}_{\rm OB}^{~2}$ term
represents the noise within an aperture of the size of the seeing disk
(assuming purely Gaussian noise) so that the weighting scheme optimizes
the S/N ratio for point sources.  We created weight maps for each field
by averaging those of the individual OBs applying the same weighting terms
(i.e., given by Eq.~\ref{Eq-weight}).  These maps, hereafter denoted
$w_{\rm F}$, give the weighted exposure time per pixel.  They reproduce
accurately the spatial variations of the background noise within each field,
as indicated by the constant pixel-to-pixel rms over the ``noise-equalized''
maps obtained by multiplying the images with the squared-root of the
associated weight maps.

The sky subtraction algorithm in DIMSUM and our additional optimization
routines have introduced small negative biases in the background visible as
orthogonal stripes at $\rm P.A. \approx 30^{\circ}$, and as dark areas around
large sources and in crowded regions.  They are due to an overestimate of the
background levels because of insufficient source masking.  The object masks
were defined from the shallower OB images, such that emission in the extended
PSF wings around sources was not fully accounted for and very faint sources
were missed.  Since this effect is systematic in all OBs, it is enhanced and
becomes apparent in the deeper field maps.  To reduce these residuals, we
rotated a copy of the field maps, made new object masks using SExtractor
(as described in \S~\ref{Sect-det}), fitted the background in the unmasked
regions with bi-cubic splines along rows and columns, rotated the fits back,
and subtracted them from the field maps.

Our aim was to produce final mosaics as uniform as possible in effective
angular resolution and optimize the S/N ratio in the overlap areas.
We determined the effective PSF of the field maps by averaging sub-images
centered on bright, isolated, unsaturated stars (normalized to the stars'
peak flux).  Table~\ref{tab-obs} lists the FWHM of the best-fit Moffat
profile to the PSFs.  For each band separately, we convolved the three
highest resolution fields to bring all fields to a common resolution.
We computed the kernels with a Lucy-Richardson deconvolution algorithm.
We quantified the accuracy of the PSF matching from the curve-of-growth
analysis of the convolved PSFs, measured from the same stars in the
smoothed field maps.  For each band, the fraction of enclosed flux agrees
to better than 1\% in circular apertures with diameters in the range
$1^{\prime\prime} - 2^{\prime\prime}$, and within 4\% for diameters between
$0\farcs 7$ ($\rm \approx 1.5 \times FWHM$ of the PSF of the smoothed field 
maps) and 6\arcsec\ (the reference aperture for photometric calibration).

To optimize the S/N ratio in the overlap areas, we applied a similar weighting
scheme as in Eq.~\ref{Eq-weight}, with however some differences.  We omitted
the FWHM-dependent term since it is identical among the PSF-matched field
maps.  We needed to account for the effective background noise levels properly.
Using the pixel-to-pixel background rms determined over the whole field images
resulted in significant global differences between the four fields in the
noise-equalized mosaics.  This indicates that the relative rms-weighted
total integration times between the fields does not reflect the differences
in effective noise.  We attribute this to the systematics discussed in
\S~\ref{Sub-red_isaac}, which affect in different proportions and with varying
severity the data of each field.  This is evident in the results of the noise
analysis described in \S~\ref{Sub-depths}, where the deviations from the
behaviour for (uncorrelated) pure Gaussian noise of the real background rms
as a function of aperture size is noticeably different between the fields.
For the inverse variance term of the weights, we used the effective noise
derived from this analysis for an aperture diameter of $\rm 1.5 \times FWHM$.
This led to the most uniform noise-equalized mosaics in terms of background
pixel-to-pixel rms, and again optimizes for point sources.

We created three sets of weight maps for the mosaics, all based on
the fields' weight maps $w_{\rm F}$.  We made one set by averaging
the $w_{\rm F}$ maps, applying the same weights as for the images.
This set accounts for the variations of effective background noise over the
entire mosaic relevant for photometry and source detection based on a S/N
threshold (``effective weight maps,'' with values denoted by $w_{\rm eff}$).
We made a second set weighting only by the cumulative number of frames
contributing to the flux (i.e., with only the $t_{\rm int,pix}$ and no
variance term), as record of the relative total integration time per pixel
(``exposure time weight maps,'' $w_{\rm exp}$).  For the third set, we simply
averaged the $w_{\rm F}$ maps assigning equal weights so that in the mosaic,
the maximum weight in the deepest part of all four fields is identical.
This set is appropriate for detecting sources using a surface brightness
threshold as we do in \S~\ref{Sub-det}
(``adjusted weight maps,'' $w_{\rm adj}$).
We renormalized all the weight maps for the mosaics to a maximum of unity.

\subsection{FORS1 Data Reduction}   \label{Sub-red_fors}

\subsubsection{Bias, Flat Field, Sky Subtraction, and Cosmic Rays Rejection}

The bias subtraction and flat-fielding of the optical FORS1 imaging were
carried out using pipeline reduction tasks developed for the FORS instruments
\footnote{Details of the pipeline reduction for FORS1$+$2 can be found at
\hfill \\
http://www.eso.org/observing/dfo/quality/FORS/pipeline/pipe\_reduc.html}
within the ESO-MIDAS system \citep{War91}.
Bias subtraction was performed using the nightly master bias frame scaled to
the average level of the science frame in the pre- and over-scan regions.
Flat-fielding was performed with the nightly master sky flat image obtained
from series of exposures taken during twilight.

We further reduced the science frames within IRAF as follows.  We subtracted
a constant sky level corresponding to the mode of each frame.  In one $U$-band
OB, there was a residual offset in background level between the lower and upper
half of the array, which we eliminated by adding to the top half the difference
in median over 20 rows below and above the mid-array row 1024.  We identified
cosmic rays with L.A.Cosmic \citep{Dok01} and negative bad pixels by applying
a $3\,\sigma$ threshold below zero.  We updated the resulting masks to flag
out additional artefacts identified by visual inspection (e.g., occasional
bad rows and columns, transient positive source-like features and broad faint
tracks, edges of the low sensitivity ``holes'' intrinsic to the FORS1 array
not adequately flat-fielded out).

We removed residual large-scale patterns (at a few percents of the
subtracted constant sky level) using a background map generated with the
SExtractor program \citep{Ber96} run on the initial sky-subtracted frames.
We specified a grid mesh size of 64 pixels and median filter size of 10
meshes for the background estimation, which produced the best results.
It is unclear whether these residuals were due to flat field inaccuracies
or spatial structure in the background.  Either division by or subtraction
of the SExtractor background map reduced the residuals to $\leq 1\%$.  The
two methods led to essentially indistinguishable results, with differences
in remaining gradients well below 1\% and in photometry of several stars in
the field smaller than $\rm 0.002~mag$.  As reported on the ESO FORS1$+$2 Web
pages, flat fielding with the master sky flat can leave large-scale gradients
of a few percents, so we adopted the sky-subtracted frames divided by the
SExtractor background maps.

\subsubsection{Flux Calibration}

We determined the photometric ZPs from observations of standard stars taken
during the same nights as the \mscl\ imaging.  These were available for the
single OB in the $V$ band, and two OBs in each the $B$ and $U$ bands.  The
stars were selected from the catalogue of \citet{Lan92}.  Several exposures
were obtained through each night, and each exposure includes multiple standard
stars.  The stars data were bias-subtracted and flat-fielded in the same way
as the \mscl\ data.  We determined single-star ZPs from individual star counts,
and accounted for the non-negligible effects of airmass and colour terms using
the atmospheric extinction and colour coefficients for the corresponding period
of observations reported on the ESO/FORS1 Web pages.  We computed the nightly
ZPs by averaging with 2$\sigma$-clipping the single-star ZPs.  The $1\sigma$
dispersions range from 0.013 to 0.049~mag, depending on the night and band. 
We also computed single-frame ZPs as the 2$\sigma$-clipped average of the
multiple stars ZPs per frame.  Frame-to-frame variations in ZPs are smaller
than the single-star ZPs dispersion and there is no trend with time, indicating
that the nights were photometric.  As for the ISAAC data, we used a circular
aperture of diameter 6\arcsec\ for the photometric calibration.  We verified
that no measurable systematic variations with airmass or colours were left.
The adopted ZPs as well as the extinction and colour coefficients used are
given in Table~\ref{tab-zps}.

For the \mscl\ data collected on nights without appropriate standard stars
observations (one OB in $B$ and three OBs in $U$), we proceeded as for the
ISAAC data taken during non-photometric nights.  We first determined the
magnitudes of five bright, isolated, unsaturated stars from the subset of
calibrated \mscl\ frames, averaging with $2\sigma$ clipping the individual
measurements (the mean of the $1\sigma$ dispersions are 0.013~mag in $B$
and 0.036~mag in $U$).  We then computed for each star and frame the
appropriate scaling factor based on the stars' magnitudes and adopted ZPs.
We calibrated the frames individually using the $2\sigma$-clipped average
of the scaling factors of all stars per frame.  We estimate the internal
accuracy of the photometric calibration of the FORS1 optical \mscl\ data
set to be 0.015~mag in $V$, 0.02~mag in $B$, and 0.04~mag in $U$.

\subsubsection{Combination and Rectification}

We aligned all calibrated frames allowing for integer-pixel shifts.  We
then co-averaged all registered calibrated frames for each band, applying the
weighting scheme of Eq.~\ref{Eq-weight} with the various terms determined from
the individual frames.  This was done before rectification to the WFPC2 maps.
Because of the smaller number of frames and the poorer seeing conditions for
the FORS1 observations (compared to the ISAAC observations), the smearing due
to the combination of dithering, uncorrected field distortion, and sub-pixel
misalignment are not critical.  The co-averaging, rectification, and creation
of the weight maps were done in the exact same way as for the ISAAC field maps.

\section{FINAL IMAGES}    \label{Sect-images}

\subsection{General Properties}   \label{Sub-improp}

The final reduced images have a pixel size of $0\farcs 1$ and are aligned
relative to the WFPC2 $I_{814}$ mosaic, with North oriented nearly up at
$0\fdg 78$ anticlockwise from the vertical axis (see \S~\ref{Sub-astrometry}
for the exact astrometry).  The images are all normalized to instrumental
counts per second.
The fully reduced images as well as the raw data are available electronically
on the FIRES Web site\footnote{http://www.strw.leidenuniv.nl/\~{ }fires}.

The ISAAC NIR mosaics and FORS1 optical maps are shallower towards the outer
edges, reflecting lower total exposure times because of the dithering process
in the observations.  In the NIR mosaics, the largest weights are reached in
the regions where the adjacent fields overlap (12\% of the total area mapped
with ISAAC).  In addition, the combination of different integration times and
noise levels results in significant global offsets in weight between the four
fields.  The largest effective weights in the central parts of the fields are
$w_{\rm eff} \approx 0.9$ for all NIR bands (in F1 for $J_{\rm s}$ and
$K_{\rm s}$, and F4 for $H$) compared to the normalized maximum of 1
reached in the overlap regions.  Field F2 has the lowest maximum effective
weight, reaching $w_{\rm eff} = 0.64$, 0.5, and 0.39 in $J_{\rm s}$, $H$, and
$K_{\rm s}$, respectively.  This can bias area selection when applying weight
criteria.  To take advantage of the field size, it may be more appropriate to
use the adjusted weight maps where the maximum weights are equal in all fields
(see \S~\ref{Sub-red_isaac}) or to treat the fields individually using their
respective weight maps $w_{\rm F}$.  The area in the $K_{\rm s}$-band mosaic
with $w_{\rm adj} \geq 0.8$, 0.25, and 0.01 (corresponding approximately to
$w_{\rm F} \geq 0.95$, 0.30, and 0.01 in at least one of the four ISAAC fields)
is 16.3, 26.2, and $\rm 28.7~arcmin^{2}$.  As quality cut for photometry,
we adopt the $\rm 23.6~arcmin^{2}$ area defined by $w_{\rm adj} \geq 0.25$
(approximately $w_{\rm F} \geq 0.3$ in at least one ISAAC field) in each
NIR band, and $w \geq 0.3$ in each optical band.
 
The noise-equalized $K_{\rm s}$-band mosaic (i.e. the $K_{\rm s}$-band map
multiplied by the squared root of its $w_{\rm eff}$ weight map) is shown in
Figure~\ref{fig-fovs}.  It illustrates the richness in faint sources and the
spatial uniformity in normalized background rms achieved with the effective
weight map.  Figure~\ref{fig-IJK} shows the RGB colour composite image of
the central $4^{\prime} \times 4^{\prime}$ of the field, constructed from
the $I_{814}$, $J_{\rm s}$, and $K_{\rm s}$ maps.  The three maps have been
convolved to an angular resolution of $\rm FWHM = 0\farcs 69$ to match the
broadest PSF of all images (in the $U$ band; see \S~\ref{Sub-imquality} below).
The colour image is displayed with a linear stretch adjusted to enhance the
fainter objects.  The variety of colours is clearly visible over a wide range
of brightnesses, even at the faintest levels.  This indicates that the maps
reach comparable depths, and reflects the combination of different rest-frame
colours and different redshifts of the sources.

\subsection{Image Quality and PSF-Matching}   \label{Sub-imquality}

We characterized the PSF in the final combined maps using bright, isolated,
and unsaturated stars (the sets were partly different in the various bands
and included eight to thirteen stars).
The PSF in the ISAAC and FORS1 images is stable and symmetric
over the entire field of view, with a Gaussian core profile and
wings consistent with a Lorentzian profile.  We adopted the average
of the peak-normalized star profiles as the final PSF in each band.
From Moffat profile fits, the PSF FWHM in the $J_{\rm s}$-, $H$-, and
$K_{\rm s}$-band mosaics is $0\farcs 48$, $0\farcs 46$, and $0\farcs 52$,
with ellipticities of 0.02, 0.04, and 0.03, respectively.  For the optical
bands, the PSF FWHMs and ellipticities are $0\farcs 69$ and 0.06 in $U$,
$0\farcs 57$ and 0.04 in $B$, $0\farcs 65$ and 0.02 in $V$, $0\farcs 21$
and 0.02 in $V_{606}$, and $0\farcs 22$ and 0.02 in $I_{814}$.  The FWHM of
the individual stars are within $\approx 10\%$ of the final PSF FWHM in each
of the ISAAC and FORS1 images, and within $\approx 30\%$ for the WFPC2 images.
The PSF FWHMs are reported in Table~\ref{tab-obs}.

For consistent photometry across all bands, we convolved all images to a
common angular resolution corresponding to that of the $U$-band map, which
has the poorest resolution ($\rm FWHM = 0\farcs 69$).  The convolution kernels
between the $U$-band PSF and that of the other bands were computed using a
Lucy-Richardson deconvolution algorithm.
From the curve-of-growth analysis of the convolved PSFs, the fraction of
enclosed flux agrees to $\leq 2\%$ in circular apertures with diameters
between 1\arcsec\ and 2\arcsec, relevant for the colour measurements
(see \S~\ref{Sub-phot}).  The accuracy of the PSF-matching is $\leq 4\%$
for diameters in the range $1^{\prime\prime} - 6^{\prime\prime}$.

\subsection{Astrometry}   \label{Sub-astrometry}

Accurate registration among the various maps is important for correct
cross-identification of sources, precise colour measurements, and proper
comparisons of the morphology in different wavebands.  We verified the
relative alignment using 20 bright stars and compact sources throughout
the field.  The median rms variation in position of individual objects
across the bands is $0\farcs 07$ (or 0.7~pixel in the final images), with
a range from $0\farcs 025$ to $0\farcs 20$.  Relative to the positions in
the WFPC2 $I_{814}$ or $V_{606}$ maps, there is no systematic trend in the
position offsets in any of the bands.  The amplitude of the offsets in the
$J_{\rm s}$, $H$, and $K_{\rm s}$ mosaics has a $1\sigma$ dispersion of
$\approx 0\farcs 07$, corresponding to a fraction 0.47 of a pixel at the
original pre-rectified scale of the ISAAC data ($\rm 0\farcs 148~pixel^{-1}$).
For the $U$, $B$, and $V$ maps, the offsets dispersion is $\approx 0\farcs 12$,
or 0.6~pixel at the original FORS1 scale ($0\farcs 20$).

We verified the absolute astrometry using stars from the USNO catalogue
\citep[v. A2.0,][]{Mon98} and UCAC2 catalogue \citep{Zac04}.  We solved
for the positions determined from the WFPC2 $I_{814}$ mosaic, our reference
image for the relative registration.  The astrometric solution is
\begin{eqnarray}
\Delta\,{\rm R.A.} = - {\rm x}\,s\,\cos(\varphi) - {\rm y}\,s\,\sin(\varphi) 
                     + 208\farcs 566
\nonumber \\
\Delta\,{\rm DEC} = -{\rm x}\,s\,\sin(\varphi) + {\rm y}\,s\,\cos(\varphi)
                    - 194\farcs 435
\label{Eq-astrom}
\end{eqnarray}
where $\rm x$ and $\rm y$ are the pixel coordinates in the final images,
for which the precise spatial scale is $s = 0\farcs 0996~{\rm pixel}^{-1}$. 
The orientation is such that North and East coincide nearly with the
vertical and horizontal axes, respectively, with a small rotation angle of
$\varphi = 0\fdg 77754$ counterclockwise (following the usual astronomical
convention, East is $90^{\circ}$ counterclockwise from North).  The offsets
$\Delta\,{\rm R.A.}$ and $\Delta\,{\rm DEC}$ are expressed in arcseconds
and relative to the reference coordinates
$\alpha_{2000}$: $\rm 10^{h} 57^{m} 00\fs 07$,
$\delta_{2000}$: $-03^{\circ} 37^{\prime} 36\farcs 19$.
We estimate that the absolute astrometry is accurate to $\leq 0\farcs 5$.

\subsection{Background and Limiting Depths}   \label{Sub-depths}

In the raw data, the background noise properties are expected to be
well described by the dispersion of the signal measured in each pixel
since both the Poisson and read-out noise should be uncorrelated.
For such uncorrelated Gaussian noise, the effective background
rms for an aperture of area $A$ is simply the pixel-to-pixel rms
$\overline{\sigma}$ scaled by the linear size $N = \sqrt A$ of the
aperture, $\sigma(N) = N\,\overline{\sigma}$.  Instrumental features,
the data reduction, and the rectification, combination, and PSF-matching
procedures have added significant systematics and correlated noise in our
final data.  The determination of the limiting depths and photometric
uncertainties depends critically on the accurate description of the
noise properties of the images.  For this purpose, we followed the
same empirical approach as for the HDF-S data \citep{Lab03a}.

We derived the function $\sigma(N)$ directly from the data by measuring the
fluxes of 2800 non-overlapping circular apertures.  We placed the apertures
at random in the maps but so as to avoid all pixels associated with sources
detected in the $K_{\rm s}$ band and within their
$\mu(K_{\rm s,AB}) = 24.8~{\rm mag\,arcsec^{-2}}$ isophote (the surface
brightness threshold applied for source detection; see \S~\ref{Sect-det}).
We used the same positions for all maps, and repeated the measurements with
different aperture diameters ranging from $0\farcs 7$ to 3\arcsec.  Because
the noise properties vary importantly between the four fields of the NIR
mosaics, we analyzed them separately (each field has about 700 aperture
positions).  We applied the procedure on the PSF-matched images used for
the photometry, allowing us to compute realistic photometric uncertainties.
We also ran the procedure on the set of non-concolved WFPC2 and ISAAC images
to estimate more representative intrinsic depths, which we can also better
compare with those of the HDF-S data (see below).

For each aperture size, the distribution of fluxes is symmetric around the
background value (about 0), and is well reproduced with a Gaussian profile.
We thus determined the background rms variations from the FWHM of the best-fit
Gaussian to the observed distribution for each aperture size.  While at any
fixed spatial scale, the noise properties over the images are consistent with
a pure Gaussian, the variations with $N$ deviate appreciably from Gaussian
scaling, being systematically larger and increasing non-linearly.  As for
the HDF-S data \citep{Lab03a}, the real noise behaviour in the \mscl\ data
set can be well approximated by the polynomial model:
\begin{equation}
\sigma_{i}(N) =
\frac{N\,\overline{\sigma}_{i}\,\left(a_{i} + b_{i}\,N\right)}{\sqrt w_{i}},
\label{Eq-noise}
\end{equation}
where $i$ designates the band (and the field for the ISAAC data) and the
weight term accounts for the spatial variations in noise level related to
the exposure time.  The coefficient $a$ represents the effects of correlated
noise, which are dominated by the resampling of the images to the WFPC2 pixel
scale and, for the PSF-matched data, the convolution applied.  The curvature
in the $\sigma_{i}(N)$ relations quantified by the coefficient $b$ indicates
a noise contribution that becomes increasingly important on larger scales.
It presumably originates from objects below the detection level, instrumental
features, and residuals from the sky subtraction and flat-fielding.

The noise behaviour is qualitatively the same in all of the images.
Figure~\ref{fig-noise} shows the results for the four fields of the
PSF-matched $K_{\rm s}$-band mosaic, with the background rms measurements for
the various apertures and the best-fit polynomial models (Eq.~\ref{Eq-noise})
compared to the expected linear relationship for uncorrelated Gaussian noise.
The flux distributions from which the background rms is derived are also
plotted for selected aperture sizes.  With the empty areas defined from
the $K_{\rm s}$-band map, sources that are undetected in $K_{\rm s}$ but
detectable in the other bands will contribute to the aperture flux
distributions.  This is reflected in the distributions for bands other
than $K_{\rm s}$, more obviously for the optical bands, with a small
excess of positive flux measurements.  However, the bias introduced in
the results is very small because the positive tail affects very little
the Gaussian fits.

Table~\ref{tab-noise} summarizes the results for the PSF-matched images,
along with the corresponding $1\sigma$ background-noise limiting magnitudes.
Table~\ref{tab-depth} gives the limiting magnitudes for the rectified but
unconvolved images.  The limiting magnitudes refer to a ``point-source
aperture'' with a diameter $d = 1.5 \times {\rm FWHM}$ that maximizes the
S/N of photometric measurements in unweighted circular apertures of point-like
sources.  Compared to the HDF-S data set, the \mscl\ field limiting magnitudes
are $\rm 0.5 - 1.9~mag$ brighter, depending on the band.  This comparison is
rather crude because it does not account for the partly different bandpasses
in the optical ($U$ and $B$ in particular) or for the different angular
resolution between the data sets.

For the ISAAC NIR data, the $J_{\rm s}$-, $H$-, and $K_{\rm s}$-band
PSF-matched maps are $\rm \approx 0.8~mag$ shallower than the HDF-S images.
The latter were matched to the resolution of the ISAAC $H$-band map of
$\rm FWHM = 0\farcs 48$ and the point-source aperture has $d = 0\farcs 7$
\citep{Lab03a}.  The intrinsic resolution of the \mscl\ NIR rectified but
unconvolved mosaics is similar, enabling a better comparison.  The limiting
magnitudes derived from this set are $\rm \approx 0.65~mag$ brighter than
for the HDF-S.  The differences expected based only on the integration
time are of $\rm \approx 0.9~mag$ for similar observing conditions (seeing,
atmospheric transmission, and night-sky OH and continuum emission).  The
depths achieved for the \mscl\ field exceed slightly these expectations.
Various factors can explain this performance.  The most important one is
probably that about half of the \mscl\ data were taken after the 2001 March
realuminization (period P3), when the effective sensitivity had increased
by $\approx 20\%-80\%$ compared to earlier epochs since the end of 1999
(P1 and P2) during which the HDF-S observations were carried out
\citep[\S~\ref{Sub-obs_isaac}; see also ][]{Lab03a}.

We also analyzed the distributions of flux ratios between bands.  The
results obtained for ratios measured in random pairs of apertures (i.e., not
necessarily spatially coincident) correspond closely to the predictions based
on Eq.~\ref{Eq-noise} assuming the noise is uncorrelated between bands.  In
contrast, the scaling relations derived for ratios within registered apertures
for several band pairs lie systematically below such predictions on scales of
$N \ga 10 - 15~{\rm pixels}$ (although still well above expectations for pure
uncorrelated Gaussian noise).  This suggests that the background noise pattern
is correlated between the bands, as was also seen in the HDF-S data set
\citep{Lab03a}.  The effects are most important and up to $\approx 20\%$
among the maps obtained with the same instrument.  This is likely due to
the large-scale residuals remaining after the data processing (e.g., from
the bias, sky, and flat-field), which are similar for a given instrument.
Interestingly, this effect is also present at the 10\% level for ratios 
involving the FORS1 $B$ or $V$ bands and the WFPC2 $V_{606}$ or $I_{814}$
bands, as well as between the WFPC2 $I_{814}$ and ISAAC $J_{\rm s}$ bands.
This is probably the signature of sources undetected in $K_{\rm s}$ but
clearly present in the other bands contaminating the ``empty'' background
flux measurements.  Confusion noise was argued in the case of the HDF-S
data but is unlikely for the shallower \mscl\ maps.  The implications of
this analysis is that our uncertainties on colours are probably slightly
overestimated because we propagated those of each bandpass neglecting the
cross-correlation terms.  We did not attempt to account for these effects
as they are much smaller than the difference between uncorrelated Gaussian
scaling and our empirical relationships.

\section{SOURCE DETECTION AND PHOTOMETRY}    \label{Sect-det}

Our aim was to construct a $K_{\rm s}$-band selected source catalogue as
uniform as possible over the entire \mscl\ field, optimized for point-like
sources, and with reliable detections and optical-NIR photometry.  Because
of the different noise properties between the four ISAAC fields, detection
with a constant S/N threshold would result in different limiting magnitudes
and limiting surface brightnesses from one field to the other.  We therefore
applied a constant surface brightness threshold $\mu(K_{\rm s})$ instead.
We compromised between completeness and reliability by choosing a
$\mu(K_{\rm s})$ criterion so as to keep the contamination by spurious
(faint) sources below $\sim 5\%$ in all fields.  The general procedure
for source detection and photometry is analoguous to that followed by
\citet{Lab03a} in making the HDF-S catalogue.

\subsection{Detection}         \label{Sub-det}

We carried out the source detection using the SExtractor software version
2.2.2 \citep{Ber96}.  The objects were detected in the rectified but non
PSF-matched $K_{\rm s}$-band map.  To optimize for point-like sources, we
filtered the detection map with a Gaussian kernel of $\rm FWHM = 0\farcs 52$
approximating well the core of the $K_{\rm s}$-band PSF.
The detection criterion was that at least one pixel be
brighter than $\mu(K_{\rm s,AB}) = 24.8~{\rm mag~arcsec^{-2}}$.
This corresponds to $\approx 5\sigma$ of the background noise in the
filtered $K_{\rm s}$-band mosaic over the two fields with intermediate
depth (F3 and F4).  The threshold adopted is rather conservative but
ensures that the spurious source contamination is small.

The resulting catalogue contains 1858 objects.  We estimated the fraction
of false detections by running SExtractor with the exact same parameters on
the inverse detection map; it is $< 1\%$ over the entire mosaic and $< 3\%$
in the shallowest field, F2.  SExtractor identifies and separates blended
sources based on the distribution of the $K_{\rm s}$-band light in the
detection map; 18\% of the sources were flagged as blended.
We specified deblending parameters so as to avoid oversplitting
of objects with relatively complex morphologies.

The spatial filtering affects the detection process and the isophotal
parameters.  Our filtering with a kernel reproducing the $K_{\rm s}$-band
PSF introduces a small bias against faint extended sources.  As for the
HDF-S catalogue \citep[see][]{Lab03a}, we preferred not to combine multiple
catalogues obtained with different kernel sizes because their subsequent
merging is complicated and subjective, and does not lead to a substantial
gain in sensitivity for larger objects.  The PSF-like kernel we adopted is
fully adequate for our purposes since most of the faint sources detected are
compact or unresolved in the NIR and it reduces the problems due to blending
when convolving with larger kernels.

\subsection{Photometry}    \label{Sub-phot}

We used SExtractor in dual-image mode, which allows for spatially
accurate and consistent photometry across all bands.  While the sources
were extracted and the isophotal parameters determined from the filtered
$K_{\rm s}$-band map, the photometry was performed on the PSF-matched maps
and within the same set of apertures in all bands for each object.  Of the
1858 $K_{\rm s}$-band detected sources, 1663 have full eight-band photometry.
We used circular apertures with 30 different diameters ranging from
$d = 0\farcs 7$ to 3\arcsec.  We also used isophotal apertures defined by
the detection threshold of $\mu(K_{\rm s,AB}) = 24.8~{\rm mag~arcsec^{-2}}$
and autoscaling apertures, which, following \citet{Kro80}, are elliptical
apertures scaled based on the first moments of the filtered $K_{\rm s}$-band
light profile (the ``MAG\_ISO'' and ``MAG\_AUTO'' computed by SExtractor).
Adopting the notation of \citet{Lab03a}, we will refer to these apertures
as APER($d$), APER(ISO), and APER(AUTO), respectively.  We also defined
customized ``colour'' and ``total'' apertures APER(COLOUR) and APER(TOTAL).
These were introduced for the HDF-S catalogue to allow detailed control
of the photometry relying on well-defined criteria and simple but robust
treatment of blended sources.  We considered a source as blended if the
SExtractor flags ``blended'' or ``bias'' were set \citep[see][]{Ber96}.

The colour aperture is optimized for accurate and consistent colours,
and relies on the $K_{\rm s}$-band isophotal aperture of each object.
The criterion to select the appropriate colour aperture uses the equivalent
circularized isophotal diameter $d_{\rm iso} = 2\,\sqrt{A_{\rm iso}/\pi}$.
For isolated sources,
\begin{eqnarray}
{\rm APER(COLOUR)} = \left\{
 \begin{array}{ll}
 {\rm APER(ISO)}       & \hphantom{hh} 1\farcs 0 < d_{\rm iso} < 2\farcs 0, \\
 {\rm APER(1\farcs 0)} & \hphantom{hh} d_{\rm iso} \leq 1\farcs 0, \\
 {\rm APER(2\farcs 0)} & \hphantom{hh} d_{\rm iso} \geq 2\farcs 0.
 \end{array}
\right.
\label{Eq-colour_isol}
\end{eqnarray}
For blended sources,
\begin{eqnarray}
{\rm APER(COLOUR)} = \left\{
 \begin{array}{ll}
 {\rm APER}(d_{\rm iso}/s) &
                    \hphantom{h} 1\farcs 0 < d_{\rm iso}/s < 2\farcs 0, \\
 {\rm APER(1\farcs 0)}     & \hphantom{h} d_{\rm iso}/s \leq 1\farcs 0, \\
 {\rm APER(2\farcs 0)}     & \hphantom{h} d_{\rm iso}/s \geq 2\farcs 0,
 \end{array}
\right.
\label{Eq-colour_blend}
\end{eqnarray}
where $s$ is the factor by which we reduce the size of the circular
apertures centered on blended objects to minimize flux contamination
from the close neighbour(s).  The optimal value of $s$ needs be determined
from experimentation, and we found that the same $s = 1.4$ as for HDF-S was
appropriate for the \mscl\ field data.  The smallest colour aperture allowed,
APER(1\farcs 0), has a diameter $\rm 1.5 \times FWHM$ of the PSF-matched images.
This maximizes the S/N of flux measurements for point sources in unweighted
circular apertures and avoids smaller apertures for which the photometry
becomes less reliable.  The largest colour aperture considered,
APER(2\farcs 0), was selected to avoid the uncertainties of very large
isophotal areas and to reduce possible contamination of the fluxes by
adjacent sources undetected in the $K_{\rm s}$ band but present in
other bands.

The total apertures were used to derive the integrated fluxes of the objects
in the $K_{\rm s}$-band.  Isolated and blended sources were again treated
differently:
\begin{eqnarray}
{\rm APER(TOTAL)} = \left\{
 \begin{array}{ll}
 {\rm APER(AUTO)}   & \hphantom{h} {\rm isolated~sources} \\
 {\rm APER(COLOUR)} & \hphantom{h} {\rm blended~sources}.
 \end{array}
\right.
\label{Eq-total}
\end{eqnarray}
Equivalent circularized apertures were also defined for the autoscaling
Kron-like apertures with $d_{\rm auto} = 2\,\sqrt{A_{\rm auto}/\pi}$ and
for the total apertures with $d_{\rm tot} = 2\,\sqrt{A_{\rm tot}/\pi}$.
To compute the total fluxes and magnitudes, we applied an aperture
correction to the measurements in APER(TOTAL).  The correction factor
is the ratio of the fluxes enclosed within circular apertures of diameters
6\arcsec\ (used for the photometric calibration) and $d_{\rm tot}$, inferred
from the growth curve of the $K_{\rm s}$-band PSF constructed from bright
stars (\S~\ref{Sub-imquality}).  This correction is significant for most
objects, reaching 0.58~mag for faint point-like sources with flux measured
in a $d = 1\farcs 0$ aperture.

Figure~\ref{fig-magtot_comp} compares various methods of estimating integrated
$K_{\rm s}$-band magnitudes.  For point-like sources, the total magnitude is
expected to correspond to the flux in the point-source APER(1\farcs 0) scaled
by the aperture correction.  The isophotal and autoscaling apertures tend to
decrease in size at fainter magnitudes, leading to more severe underestimate
of the total magnitudes due to extended emission intrinsic to the sources or in
the PSF wings outside of the apertures.  For isophotal magnitudes, the turnover
sets in at $K_{\rm s,AB}^{\rm tot} \approx 22.5~{\rm mag}$, with the difference
$K_{\rm s}^{\rm iso} - \left[K_{\rm s}(1\farcs 0)-0.58\right]$ growing rapidly
and reflecting mainly the shrinking APER(ISO) defined by the constant surface
brightness threshold.  Magnitudes within APER(AUTO) come closer but still miss
significant flux in the PSF wings at $K_{\rm s,AB}^{\rm tot} \ga 24~{\rm mag}$. 
With the method based on Eq.~\ref{Eq-total}, we best recover the integrated
magnitudes expected for faint point-like sources.  Our treatment of isolated
and blended sources also reduces the dispersion of $K_{\rm s}^{\rm tot}$
around $K_{\rm s}(1\farcs 0) - 0.58$.

We derived the $1\sigma$ photometric uncertainties from Eq.~\ref{Eq-noise},
using the $a_{i}$ and $b_{i}$ coefficients given in Table~\ref{tab-noise},
the linear size $N = \sqrt{A}$ of the aperture considered, and the average
weight within the aperture.  These uncertainties properly account for the
actual background noise in the maps influencing directly the flux within the
various apertures.  They may overestimate the uncertainties of colours by a
small amount (\S~\ref{Sub-depths}).  From the variations in photometric ZPs
determined from standard stars or inferred from stars within the \mscl\ field
of view (\S~\ref{Sub-red_isaac} and \S~\ref{Sub-red_fors}), we estimate that
uncertainties of the absolute flux calibration are $\rm \approx 0.05~mag$.
The photometry may further be affected by possible biases from the sky and
background subtraction procedure, or surface brightness biases; these
additional sources of errors are not included in our uncertainties.

\section{REDSHIFTS}      \label{Sect-redshifts}

For all objects in the catalogue, we derived photometric redshifts
($z_{\rm ph}$) using a technique described in detail by \citet{Rud01, Rud03}
and \citet{Lab03a}.  Briefly, the algorithm involves linear combinations
of redshifted spectral templates of galaxies of various types (from normal
ellipticals and spirals to magellanic irregulars and young starbursts).
The mean absorption by neutral hydrogen from the intergalactic medium
along the line of sight is accounted for \citep[based on][]{Mad95}.
The $z_{\rm ph}$ uncertainties are estimated from Monte-Carlo simulations,
accounting for uncertainties in the fluxes as well as template mismatch.
A key feature of the technique is that the derived uncertainties reflect
not only the $1\sigma$ confidence interval around the best solution but
also the presence of secondary solutions with comparable likelihood.

Spectroscopic redshifts ($z_{\rm sp}$) have been
measured for over 400 galaxies in the \mscl\ field 
\citep[][S. Wuyts et al., in preparation]{Tra99, Tra03, Dok03, Dok04},
of which about 330 are included in our $K_{\rm s}$-band selected catalogue.
The majority ($\approx 85\%$) lie at $z_{\rm sp} < 1$, mostly members of
the \mscl\ cluster at $z = 0.83$.  Of the remaining, 32 galaxies have
$1 < z_{\rm sp} < 2$ and 11 have $z_{\rm sp} > 2$.  The much larger number
of $z_{\rm sp}$ determinations in the \mscl\ field compared to HDF-S allows a
more robust assessment of the accuracy of our photometric redshift technique.

Figure~\ref{fig-zph_zsp} compares the $z_{\rm ph}$ and
$z_{\rm sp}$ for galaxies with spectroscopic determinations,
and shows the $z_{\rm ph}$ and $z_{\rm sp}$ distributions of
sources at $K^{\rm tot}_{\rm s,AB} \leq 24.8~{\rm mag}$,
for which ${\rm S/N}(K_{\rm s,AB}^{\rm tot}) \geq 5$.
Overall, the accuracy of the photometric redshifts is
${\rm d}z / (1 + z_{\rm sp}) \equiv 
\langle |z_{\rm ph}-z_{\rm sp}|/(1+z_{\rm sp}) \rangle = 0.078$.
In the redshift ranges $z < 1$, $1 < z < 2$, and $z > 2$,
${\rm d}z / (1 + z_{\rm sp}) = 0.074$, 0.127, and 0.040, respectively.
At $1 < z < 2$, the $z_{\rm ph}$'s are less well constrained, as reflected by
the larger $1\sigma$ confidence intervals.  This is mainly because the fitting
procedure there relies heavily on the presence of the Balmer/4000\,\AA\ break
in the SEDs (characteristic of $\rm \ga 10^{8}~yr$ old stellar populations),
and the break falls in the large gap between the $I_{814}$ and $J_{\rm s}$
bandpasses.  At $z > 2$, the agreement between $z_{\rm ph}$ and $z_{\rm sp}$
is best, although the number of galaxies with available $z_{\rm sp}$ in this
redshift interval is still small.  Our $z_{\rm ph}$'s slightly underestimate
the true redshifts by $z_{\rm ph} - z_{\rm sp} \approx 0.1$ on average.
In the rest of the paper, we will use the spectroscopic redshift whenever
available, otherwise the photometric redshift.

We identified stars using a combination of three criteria.  The main criterion
was that the raw $\chi^{2}$ for fits using stellar SEDs was lower than that of
the best-fit combination of galaxy SEDs from the $z_{\rm ph}$ determination.
The stellar templates are synthetic SEDs of main-sequence stars with effective
temperatures from 3000 to 10000~K computed with the NEXTGEN model atmospheres
of \citet{Hau99}.  We did not consider cooler dwarfs or evolved stars (giants,
supergiants) because non local thermodynamical equilibrium (non-LTE) effects
and extensive molecular absorption become important and model atmospheres still
suffer from large uncertainties in these regimes.  Moreover, adequate empirical
spectral atlases with complete and consistent optical-NIR coverage are scarce
for these stellar classes.  As argued for the HDF-S, very late M-type, methane,
T-, and L- dwarfs as well as normal post-main-sequence or variable O- to C-rich
stars on the asymptotic giant branch are not expected to be as red in the NIR
as most redshifted galaxy types and/or are very unlikely to be present in the
small survey areas at high galactic latitude covered by the HDF-S and \mscl\
fields.  As noted by \citet{Lab03a}, stars are well separated from galaxies
in the $J_{\rm s} - K_{\rm s}$ versus $I_{814} - K_{\rm s}$ diagram (see also
\S~\ref{Sect-analysis}); our second criterion was that the candidates lie on
the locus of stars in this colour space.  We then verified the size of each
candidate in the WFPC2 $I_{\rm 814}$ mosaic (non PSF-matched) and excluded
those that are spatially resolved.  The final list includes 84 stars down
to $K_{\rm s,AB}^{\rm tot} = 24.8~{\rm mag}$.

\section{CATALOGUE PARAMETERS}    \label{Sect-catalogue}

The source catalogue is available electronically on the
FIRES Web site\footnote{http://www.strw.leidenuniv.nl/\~{ }fires}.
The catalogue format and entries are similar to those of
the HDF-S catalogue \citep{Lab03a}, and described here.

{\em ID ---\/}
Running identification number in the catalogue order as reported by SExtractor.

{\em x, y ---\/}
Pixel positions of the objects in the images, based on the 
$K_{\rm s}$-band detection map.  The reference is the $4700 \times 4700$
frame of the rectified, registered maps at $\rm 0\farcs 1~pixel^{-1}$.

{\em RA, DEC ---\/}
Right ascension and declination coordinates for equinox $\rm J2000.0$
(see Eq.~\ref{Eq-astrom}).

{\em $f_{{\rm col},i}$, $\sigma_{{\rm col},i}$ ---\/}
Flux measured in the colour aperture (\S~\ref{Sub-phot}) and the associated
uncertainty derived from the noise analysis (\S~\ref{Sub-depths}) in each of
the bandpasses $i = U, B, V, V_{606}, I_{814}, J_{\rm s}, H, K_{\rm s}$.
The units are $\rm \mu Jy$.

{\em $f_{{\rm iso},i}$, $\sigma_{{\rm iso},i}$ ---\/}
Same as above for measurements in the isophotal
aperture defined by the surface brightness threshold of
$\mu(K_{\rm s,AB}) = 24.8~{\rm mag~arcsec^{-2}}$ in the
$K_{\rm s}$-band detection map (the MAG\_ISO computed by SExtractor).

{\em $f_{\rm tot}(K_{\rm s})$, $\sigma_{\rm tot}(K_{\rm s})$ ---\/}
$K_{\rm s}$-band flux in the total aperture scaled by the aperture correction
(\S~\ref{Sub-phot}) and the associated uncertainty from the noise analysis
(\S~\ref{Sub-depths}).  The units are $\rm \mu Jy$.  The aperture correction
is based on the $K_{\rm s}$-band PSF profile outside of the total aperture.

{\em $f_{\rm auto}(K_{\rm s})$, $\sigma_{\rm auto}(K_{\rm s})$ ---\/}
Same as above for measurements in the Kron-like autoscaling aperture
based on the light distribution in the $K_{\rm s}$-band detection map
(the MAG\_AUTO computed by SExtractor).

{\em $f_{{\rm ap},i,d}$, $\sigma_{{\rm ap},i,d}$ ---\/}
Flux in the bandpass $i$ and the associated uncertainty derived from the noise
analysis (\S~\ref{Sub-depths}) measured in each of the 30 circular apertures
with diameter $d = 7, 8, 9, 10, ..., 29, 30, 35, 40, 45, 50, 55, 60$ pixel
of the rectified maps ($\rm 0\farcs 1\,pixel^{-1}$).
The units are $\rm \mu Jy$.

{\em apcol ---\/}
Type of aperture used for measuring the colour flux (\S~\ref{Sub-phot}),
encoded as follows.
1: minimum circular aperture considered with diameter
$d = 1\farcs 0$ used whenever the isophotal aperture is smaller.
2: maximum circular aperture considered with diameter
$d = 2\farcs 0$ used whenever the isophotal aperture is larger.
3: isophotal aperture defined by the surface brightness threshold
of $\mu(K_{\rm s,AB}) = 24.8~{\rm mag~arcsec^{-2}}$ in the
$K_{\rm s}$-band detection map.
4: circular aperture with diameter $d = \sqrt{A_{\rm iso}/\pi} / s$
where $s = 1.4$ is the aperture reduction factor applied for blended sources.

{\em aptot ---\/}
Type of aperture used for measuring the total $K_{\rm s}$-band flux
(\S~\ref{Sub-phot}), encoded as follows.
1: Kron-like autoscaling aperture based on the light distribution
in the $K_{\rm s}$-band detection map.
2: Colour aperture used in the case of blended sources.

{\em apcorr ---\/}
Scaling factor corresponding to the aperture correction applied to the 
$K_{\rm s}$-band flux measured in the total aperture in order to estimate
the total integrated flux.

{\em $r_{\rm col}$, $r_{\rm tot}$ ---\/}
Circularized radii $r = \sqrt{A/\pi}$ from the area of the
apertures used to measure the colour and total fluxes.
Units are pixels in the rectified images ($\rm 0\farcs 1\,pixel^{-1}$).

{\em $A_{\rm iso}$, $A_{\rm auto}$ ---\/}
Area of the aperture enclosed within the 
$\mu(K_{\rm s,AB}) = 24.8~{\rm mag~arcsec^{-2}}$ isophote, and
of the elliptical Kron-like autoscaling aperture ($\pi\,a\,b$,
where $a$ and $b$ are the semi-major and semi-minor axis lengths).
Units are squared pixels in the rectified images ($\rm 0\farcs 1\,pixel^{-1}$).

{\em ${\rm FWHM}(K_{\rm s})$, ${\rm FWHM}(I_{814})$ ---\/}
FWHM of sources from the $K_{\rm s}$-band detection map assuming
a Gaussian profile, and from the non PSF-matched WFPC2 $I_{814}$ mosaic.
The latter was obtained by running SExtractor on the $I_{814}$ map,
and cross-correlating the $I_{814}$- and $K_{\rm s}$-band selected catalogues.
Units are pixels in the rectified images ($\rm 0\farcs 1\,pixel^{-1}$).

{\em $w_{i}$ ---\/}
Average of the effective weight in the bandpass $i$ within the
point-source aperture for the PSF-matched set of maps (diameter
of $d = 1.5 \times {\rm FWHM} = 1 \farcs 0$).  For the optical FORS1
and WFPC2 data, the effective weight is simply proportional to the total
exposure time at each pixel while for the NIR ISAAC mosaics, it corresponds
to the inverse variance of the effective background noise resulting from the
integration time and the systematics from the raw data and the reduction
procedure (\S\S~\ref{Sub-red_isaac} and \ref{Sub-red_fors}).  The weights
are normalized to a maximum of unity over the maps.

{\em $w_{F}(J_{\rm s},H,K_{\rm s})$ ---\/}
Same as above for the weights in each of the fields $F = 1, 2, 3, 4$ of the
NIR ISAAC data separately.  These weights are simply proportional to the
total exposure time per pixel, and normalized to a maximum of unity within
each field map.

{\em bias, blended, star ---\/}
Flags indicating the following if set to 1 (and otherwise set to 0).
{\em Bias\/} is set if the Kron-like autoscaling aperture APER(AUTO) contains
more than 10\% of bad pixels or if the flux measurement within this aperture
is affected by neighbouring sources as determined from the light distribution
in the $K_{\rm s}$-band detection map.
{\em Blended\/} is set if the source overlaps with adjacent objects based on
the $K_{\rm s}$-band detection map.
{\em Star\/} is set if the star identification criteria are satisfied
(SED better fit by a stellar template than a combination of galaxy templates,
colours lying on the stellar locus in the $J_{\rm s} - K_{\rm s}$ versus
$I_{814} - K_{\rm s}$ diagram, spatially unresolved light profile in the
non PSF-matched WFPC2 mosaics; \S~\ref{Sect-redshifts}).

{\em $z_{\rm ph}$, $\sigma_{\rm up,low}(z_{\rm ph})$ ---\/}
Photometric redshift and associated 68\% confidence intervals
(${z_{\rm ph}}\,^{+\sigma_{\rm up}}_{-\sigma_{\rm low}}$;
\S~\ref{Sect-redshifts}).

\section{ANALYSIS}    \label{Sect-analysis}

\subsection{Completeness in the $K_{\rm s}$ band}
            \label{Sub-complete}

We derived the completeness limits in the $K_{\rm s}$ band from
simulations in which we added point sources to the mosaic and analyzed
the recovery fraction as a function of input magnitude.  We performed
the simulations on the rectified ($\rm 0\farcs 1~pixel^{-1}$) and non
PSF-matched ($\rm FWHM = 0\farcs 52$) image used for the source detection.
We treated the four ISAAC fields separately because of their different
effective depths.  We considered only the deepest part of each field with
nearly uniform image quality, where the effective weights exceed 95\% of
the maximum within the given field (i.e., $w_{\rm F} > 0.95$).  We used the
PSF profile (created by averaging the images of bright, isolated, unsaturated
stars, see \S~\ref{Sub-imquality}) to generate artificial point sources with
total $K_{\rm s,AB}$-band magnitudes between $\approx 21$ and 28.
For realistic simulations, we distributed the sources according to
the power-law ${\rm d}N/{\rm d}K_{\rm s} \propto K_{\rm s}^{\alpha}$.
We adopted $\alpha = 0.20$ as determined from the raw number counts at
$21.4 \la K_{\rm s,AB}^{\rm tot} \la 23.4~{\rm mag}$, where $\rm S/N \ga 10$
and where incompleteness does not yet play a significant role.

For each field, we added 30000 simulated point sources to the image, drawn
at random from the power-law distribution and placed at random locations.
To avoid overlap among the simulated sources, we ran the simulations by
adding only 25 of them at a time to the image.  For each realization,
we extracted the sources as described in \S~\ref{Sub-det} except for
the application of a fainter surface brightness detection threshold of
$\mu(K_{\rm s,AB}) = 25.3~{\rm mag~arcsec^{-2}}$ ($\approx 3\sigma$ of
the background noise in the two intermediate-depth fields) to ensure
reliable derivation of the 50\% completeness limits.
We considered a source as recovered if its measured position lay
within 3.5 pixels of its input position, although experimentation showed
that the results are little sensitive to the exact choice of this criterion.
We accounted for blending with real sources following Eq.~\ref{Eq-total}
when determining the total $K_{\rm s}$-band magnitudes of the simulated
sources.  We did not apply any particular treatment for simulated sources
that may have been lost on or falsely recovered as real sources (``unmasked
case'').  To quantify this effect, we repeated the analysis excluding all
regions within the isophotal areas of real sources (``masked case'').

Figure~\ref{fig-complete} shows the completeness curves for each field and
the output versus input total $K_{\rm s}$-band magnitudes for the unmasked
case.  Table~\ref{tab-complete} gives the 90\% and 50\% completeness limits
for both the unmasked and masked cases.  For the unmasked case, the 90\%
completeness levels vary by up to 0.8~mag between the fields, with a mean
of $K_{\rm s,AB}^{\rm tot} \approx 24.1~{\rm mag}$.  The 50\% completeness
levels are essentially the same for the four fields at
$K_{\rm s,AB}^{\rm tot} = 25.30 \pm 0.01~{\rm mag}$
(and close to the $3\sigma$ detection limits).
The shape and slope of the completeness curve for the shallowest field F2 at
faint magnitudes is distinctly different from that of the other three fields,
which reach fainter and more similar depths.  This is caused by a more
important contribution from spurious detections in F2 because of higher
amplitude noise fluctuations.

Not only the depth of the images but also the crowding of real sources
influences the derived limits, through loss and confusion of simulated
sources overlapping with real objects.  This effect is significant in the
\mscl\ field compared to the HDF-S because of the presence of the cluster.
We found this alters mainly the absolute level and slope of the completeness
curves above 90\%, and therefore the inferred 90\% limits.  For comparison,
in the masked case the bright end of the curves is flatter, and the 90\%
completeness levels increase to $\rm \approx 24.7~mag$ and reflect more
closely the relative depth between the fields as derived from the background
noise analysis in \S~\ref{Sub-depths}.  In contrast, the steepest part of
the curves is more robust against loss and confusion of simulated sources
with real sources, with derived 50\% limits increasing marginally to
$\rm 25.34 \pm 0.02~mag$.

Our assumption of point-like simulated sources optimizes the detection
efficiency, which is further enhanced by our filtering of the detection map
with a PSF-like kernel (see \S~\ref{Sub-det}).  Consequently, the completeness
levels derived above constitute upper limits.  In reality, extended sources
would have brighter completeness levels that would also depend on the actual
sizes and light profiles.  An analysis as a function of source size and
morphology is presented by \citet{Tru05}.
In correcting the raw counts below, we followed the simple conservative
approach based on the point-source simulations presented above, and used
the results for the unmasked case.

\subsection{$K_{\rm s}$-band number counts}
            \label{Sub-numcounts}

We computed the raw $K_{\rm s}$-band source counts over the same area
of the mosaic as considered for the simulations presented above (where
$w_{\rm F} > 0.95$ for each individual field).
We excluded stars, used 0.5~mag wide bins, and extended the counts to the
50\% completeness level ($K_{\rm s,AB}^{\rm tot} \approx 25.3~{\rm mag}$).
In correcting the raw counts for incompleteness and spurious detections,
we accounted for the non-uniform depth and noise properties over the mosaic
as follows.  We first treated each field separately, computing the raw counts,
applying the corresponding corrections, and weighting by the effective area
where the depth equals the faint edge of each magnitude bin.  We then combined
the corrected counts from the four fields and normalized by the total effective
area over the mosaic as a function of depth to obtain the surface densities.
We derived the effective area to a given depth based on the mosaic's effective
weight map ($w_{\rm eff}$; \S~\ref{Sub-red_isaac}), which accounts for the
variations in absolute noise levels over the entire field of view.

This procedure yields the most reliable results for our \mscl\ data,
especially because of the significant offset in central depth of F2 compared
to the other fields.  Accounting for the effective area as a function of depth
has little impact within each field because we only considered the deepest
regions within each of them.  However, it becomes important when combining
the fields' counts.  In particular, at the faintest magnitude bin considered
(centered at $K_{\rm s,AB}^{\rm tot} = 25.0~{\rm mag}$), the raw counts as
well as the point-source simulations in F2 suffer from a larger contamination
by spurious detections so that the derived corrections are less reliable.
Our treating the fields separately and accounting for the effective area
per magnitude interval avoids a strong bias from F2 at the faint end.

Figure~\ref{fig-numcounts1} shows the differential $K_{\rm s}$-band galaxy
counts for the \mscl\ field.  The raw counts are plotted along with the
counts corrected for incompleteness only and for both incompleteness and
spurious detections.  The incompleteness correction is simply the inverse
of the detection efficiency of simulated sources given by the completeness
curves.  The correction for incompleteness and spurious detections is taken
as the ratio of the number of simulated sources per input magnitude bin
to the number of detections per output magnitude bin.  Since spurious
sources contribute to the number of detections (as for the raw counts),
the full correction is effectively smaller than that for incompleteness
only.  As the figure shows, little correction is needed out to the faintest
magnitude, where the incompleteness and full corrections amount to 30\% and
10\%, respectively.

The presence of the $z = 0.83$ cluster has minimal effect on the counts at
the faint end.  The average magnification factor from gravitational lensing
over the entire mosaic is small and ranges from 1.05 to 1.25 for background
source redshifts between $z = 1$ and 4 (H. Hoekstra, priv. comm.; based on
\citealt{Hoe00}).  Moreover, the flux and area magnification cancel out
to first order by shifting the counts in the surface densities$-$magnitude
plane along a path that is roughly parallel to the observed distribution
(as indicated in Figure~\ref{fig-numcounts1} for $z = 2.5$ with average
magnification factor of 1.20).
At brighter magnitudes, the cluster galaxies produce the bump around
$K_{\rm s,AB}^{\rm tot} \approx 20.5~{\rm mag}$ (the approximate cluster
contribution is illustrated in Figure~\ref{fig-numcounts1}, taken as the
counts of all sources with $0.6 < z < 1.0$, which are dominated by the
cluster galaxies; see Figure~\ref{fig-zph_zsp}).
Accurate corrections for the lensing effects and direct contribution to the
counts of the cluster are beyond the scope of this paper (and would depend,
e.g., on the luminosity function of the background sources, which is unknown,
and on a reliable and complete census of the cluster members).

Figure~\ref{fig-numcounts2} compares the FIRES counts
with those of selected deep surveys from the literature.
In the range $K_{\rm s,AB}^{\rm tot} \approx 22 - 25~{\rm mag}$
($K_{\rm s,Vega}^{\rm tot} \approx 20 - 23~{\rm mag}$), the \mscl\ field
counts follow a ${\rm d}\log(N) / {\rm d\,mag}$ relation with logarithmic
slope $\alpha = 0.20$ (it varies from 0.19 to 0.21 between the raw and
incompleteness-corrected cases).  This is comparable to the faint-end
slope derived from our $\approx 0.7~{\rm mag}$ deeper $K_{\rm s}$-band
map of HDF-S \citep{Lab03a}.
Overall, the counts in the two FIRES fields agree quite well out to the
50\% completeness limit of the shallower \mscl\ field, with the exception
of the bump due to the \mscl\ cluster itself.  At the faint end, the good
agreement suggests that faint cluster galaxies probably do not significantly
affect the count slope in the \mscl\ field.  The $\approx 5$ times wider
area of the \mscl\ field also leads to more robust statistics compared to
the deeper but smaller HDF-S.  Compared to other $K$-band surveys, the
\mscl\ field is not the deepest nor the widest, but is unique in its
combination of depth and area surveyed.

There are appreciably large differences among the various counts
shown in Figure~\ref{fig-numcounts2}.  In particular, the faint-end
slope of the FIRES fields generally lies at the flatter end of the
range from the other surveys ($\approx 0.23 - 0.36$
at $22 \la K_{\rm s,AB} \la 25-26~{\rm mag}$,
or $20 \la K_{\rm s,Vega} \la 23-24~{\rm mag}$;
\citealt*{Djo95, Mou97, Ber98, Sar99, Mai01}).
It seems possible that the \citet{Ber98} counts may have been overestimated.
The significance of the differences between studies is however difficult to
ascertain in the faint regime, where model-dependent corrections become
important.  A complex combination of effects is probably at play, including
intrinsic field-to-field variations (which can be large since all the
fields are relatively small, covering a few arcminutes squared to a few
tens of arcminutes squared) as well as differences in the photometric
measurements (e.g., estimates of total magnitudes), the filter bandpasses
($K$, $K_{\rm s}$, $K^{\prime}$), the derivations and application of the
corrections to the counts (spurious sources contaminations, non-uniform
depth, size and morphology dependencies).  This makes any comparison
between surveys far from straightforward.

\subsection{Colour and magnitude distributions}   \label{Sub-colmag}

Figures~\ref{fig-colmag} and \ref{fig-colcol} show the distribution
of $K_{\rm s}$-band selected sources from the \mscl\ field in various
$I_{814}\,J_{\rm s}\,H\,K_{\rm s}$ colour-magnitude and colour-colour
diagrams.  The sources are selected to have a $\rm S/N > 10$ on their
$K_{\rm s}$-band colour flux, and to lie within the well-exposed areas
with $w_{\rm F} \geq 0.3$ in each NIR band and $w_{\rm eff} \geq 0.3$
in each optical map (implying $K_{\rm s,AB}^{\rm tot} \la 24.4$ or
$K_{\rm s,Vega}^{\rm tot} \la 22.5~{\rm mag}$ for the resulting sample).
All colours refer to measurements in the colour apertures, as described
in \S~\ref{Sub-phot}.  The colour distributions are unaffected by the
(achromatic) gravitational lensing of background sources from the $z = 0.83$
cluster.  The cluster sequence is seen in all plots, in the tight relationships
at brighter magnitudes.  The stars identified in \S~\ref{Sect-redshifts}
follow a well-defined blue locus in
$J_{\rm s} - K_{\rm s}$ and $H - K_{\rm s}$ versus $K_{\rm s}^{\rm tot}$,
and in $J_{\rm s} - K_{\rm s}$ versus $I_{814} - K_{\rm s}$ and
$H - K_{\rm s}$ colours, well separated from the distributions for galaxies.

As for the HDF-S \citep{Lab03a}, we find large numbers of sources with
very red $J_{\rm s} - K_{\rm s}$ colours in the \mscl\ field.  In particular,
the criterion $J_{\rm s,Vega} - K_{\rm s,Vega} > 2.3~{\rm mag}$ effectively
selects evolved or dust-obscured galaxies at $z > 2$, the ``Distant Red
Galaxies'' \citep[DRGs,][]{Fra03, Dok03, Dok04}.  Their average rest-frame
optical colours fall within the range covered by normal galaxies locally,
unlike the Lyman-break galaxies (LBGs) selected in the optical as $U$-dropouts,
which are typically much bluer (\citealt{FS04}; see also \citealt{Pap01}).
This indicates that DRGs have on average higher mass-to-light ratios than LBGs
at similar redshifts.  The surface density of DRGs is $1.6 \pm 0.3$ and
$\rm 2.9 \pm 0.8~arcmin^{-2}$ in the \mscl\ and HDF-S fields, respectively,
at $K_{\rm s,Vega}^{\rm tot} < 22.5~{\rm mag}$, and their contribution
to the integrated rest-frame $V$-band luminosity is $\sim 30\%$.
This suggests that DRGs may be a significant constituent
of the $z \sim 2 - 3$ universe in terms of stellar mass
(\citealt{Fra03, FS04}; Rudnick \etal, in preparation).
Field-to-field variations are likely large for DRGs, as suggested by
the high clustering amplitude derived for the HDF-S sample \citep{Dad03}
and by the fact that DRGs lie at the high-mass end of galaxies at $z > 2$
\citep{FS04}.  These variations may explain why such red $J - K$ selected
galaxies have not been recognized as an important high redshift population
in previous NIR surveys; those deep enough to detect very red and faint
high redshift sources were limited to very small areas.

Another class of red objects, the Extremely Red Objects (EROs) at
$1 \la z \la 2$ selected by their very red $I - K$ or $R - K$ colours,
has been well-studied in the past decade \citep[see][for a review]{McC04}.
Figure~\ref{fig-colcol} shows that at
$K_{\rm s,Vega}^{\rm tot} < 22.5~{\rm mag}$
in the \mscl\ field, there is some overlap between samples selected with
the DRG criterion $J_{\rm s,Vega} - K_{\rm s,Vega} > 2.3$ and with the
ERO criterion $I_{\rm 814,Vega} - K_{\rm s,Vega} > 4$.
\footnote{Colour cutoffs to select EROs vary somewhat in the literature
depending on the filter bandpasses involved and the scientific purposes
of the authors.}
About 60\% of $J_{\rm s,Vega} - K_{\rm s,Vega} > 2.3$ objects also have
$I_{\rm 814,Vega} - K_{\rm s,Vega} > 4$ colours.  However, $\approx 30\%$
only of $I_{\rm 814,Vega} - K_{\rm s,Vega} > 4$ objects meet the DRG criterion.
This reflects primarily a redshift effect: the DRG criterion is efficient
at isolating evolved or very dusty systems at $z > 2$, as it is intended
to, while the ERO criterion is more sensitive to galaxies at lower
redshifts, down to $z \sim 1$.  Figure~\ref{fig-zhist} compares the
redshift distributions of the samples in the \mscl\ field.

\section{SUMMARY}  \label{Sect-conclu}

We have presented the observations, reduction, source detection and
photometry of the field around the \mscl\ cluster observed as part of
FIRES.  We have described the main properties of the final images and
of the $K_{\rm s}$-selected source catalogue.  The \mscl\ field complements
our HDF-S data, covering an area $\approx 5$ times wider with very similar
image quality in the NIR and reaching limiting total magnitudes for point
sources about 0.7~mag brighter.  We have compared the photometric redshifts
with the extensive set of spectroscopic redshifts available for the \mscl\
field and found very good agreement with
$\langle |z_{\rm sp} - z_{\rm ph}| / (1 + z_{\rm sp}) \rangle = 0.078$.
The \mscl\ field confirms our initial findings from the HDF-S,
with large numbers of red NIR-selected high redshift objects,
and substantially increases the statistical significance of the
samples, in particular for the Distant Red Galaxies at $z > 2$
selected by their $J_{\rm s,Vega} - K_{\rm s,Vega} > 2.3$
colours.

This paper completes the initial phase of the FIRES survey.
Figure~\ref{fig-surveys} places FIRES in a broader context by
comparing it in the area-depth plane to other published deep NIR surveys.
The HDF-S data constitute the deepest ground-based imaging at NIR wavelengths
to date.  In $J$ and $H$ band, it is only surpassed in depth by space-based
imaging with HST NICMOS while in $K$-band it represents the currently deepest
observations even from space.  The \mscl\ field occupies a unique position
in terms of area-depth combination, being the widest for its depth or
conversely the deepest for its size.

We have exploited our deep optical-NIR imaging of the HDF-S and
\mscl\ fields in several papers to investigate the properties of
the $K_{\rm s}$-selected galaxies and various sub-samples, focusing
on their SEDs, their rest-frame optical luminosities, colours, sizes,
their stellar masses, their clustering, and the derived redshift
evolution of both their integrated light and stellar mass
\citep{Fra03, Rud03, Dad03, Lab03b, Dok03, Dok04, FS04, Tru03, Tru05}.
The results of this initial phase have guided the directions of our current
and planned follow-up programs.  The scientific inferences from the HDF-S
and \mscl\ fields remain yet limited by the effects of field-to-field
variations, and the size and low spectroscopic completeness of the
high-redshift samples in particular.  We have thus actively undertaken
the addition of more disjoint fields and extensive spectroscopy.
Current and planned follow-ups also include near-infrared integral field
spectroscopy, extension of the imaging to other wavebands (especially
in the mid-IR with Spitzer), and HST ACS and NICMOS high angular
resolution imaging to constrain the nature and detailed spectral,
dynamical, and morphological properties of the galaxy samples (e.g.,
\citealt{Rub04, Lab05, Dok05, Tof05}; K. K. Knudsen \etal, in preparation).
With its unique data set, FIRES has been influential to our understanding
of galaxy evolution, and will continue to be so as it is extended by its
various follow-up programs.  It sets an important basis in the framework
of major current and planned surveys such as, for instance, GOODS and
COSMOS.

\acknowledgments
We would like to thank the ESO staff for obtaining the high quality ISAAC
and FORS1 data in service mode and for making them readily available to us.
We also thank Henk Hoekstra for kindly computing for us the lensing
magnifications in the \mscl\ field.
N.M.F.S. and G.R. acknowledge generous travel support from the
Leids Kerkhoven-Bosscha Fonds.  G.R. acknowledges the financial support
of Sonderforschungsbereich 375 from the Deutsche Forschungsgemeinschaft.
G.D.I. acknowledges financial support by the NASA grant NAG~5-7697.





\setcounter{figure}{0}

\clearpage

\begin{figure}[p]
\figurenum{1}
\epsscale{1.0}
\plotone{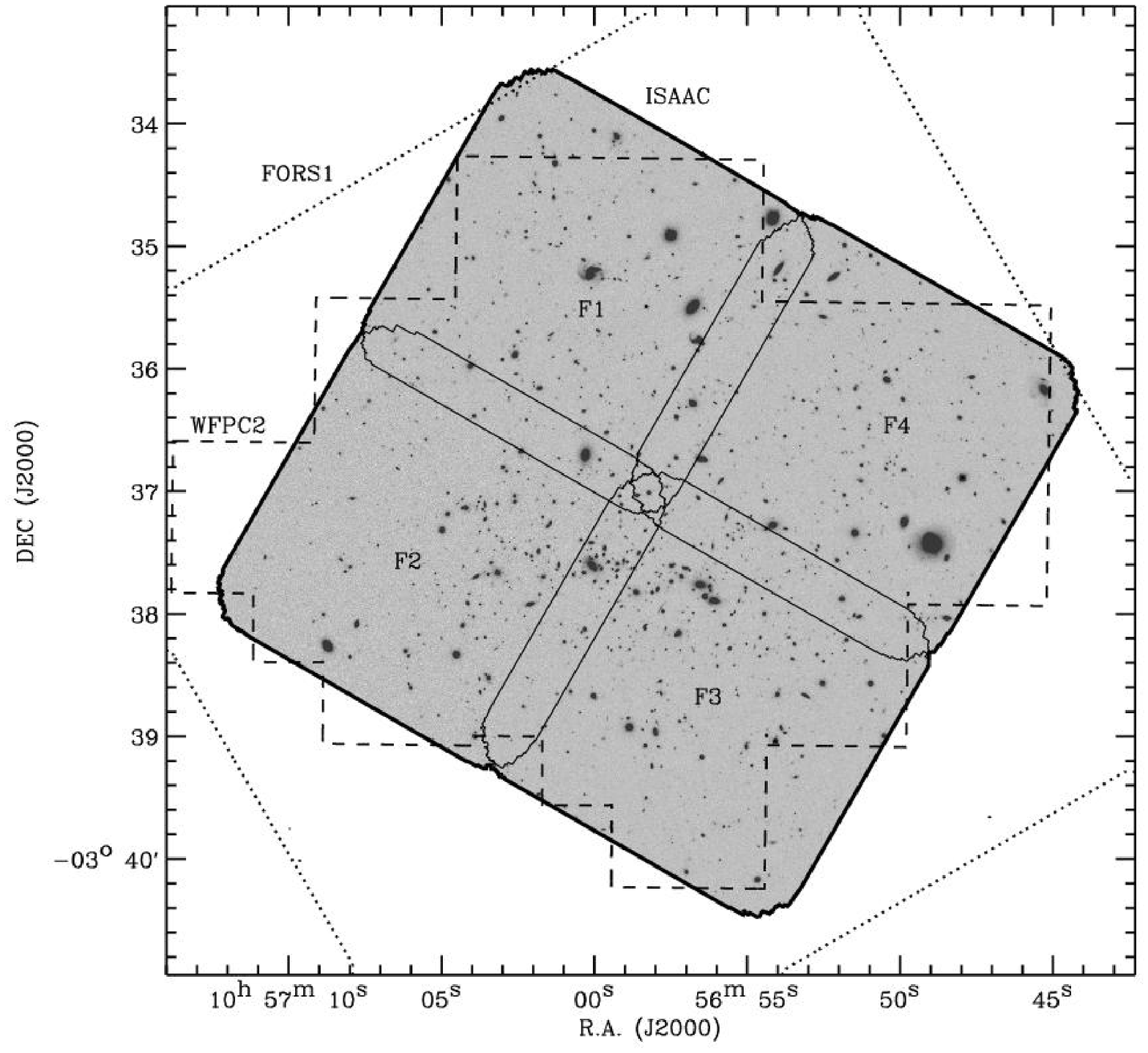}
\vspace{0.0cm}
\caption{
Fields of view of $\rm MS\,1054-03$ covered by the different instruments.
The full ISAAC field of view is outlined by the thick solid contour around
the $K_{\rm s}$-band mosaic.  The mosaic is normalized by the squared root
of the weight map (\S~\ref{Sub-red_isaac}) and displayed with a linear
intensity scaling.
The thin solid lines delineate the four individual ISAAC fields labeled F1
to F4, and show the regions of overlap.  The dashed and dotted lines indicate
the outer limits of the WFPC2 mosaic and the FORS1 field of view, respectively.
The cluster is apparent as the east-west elongated S-shaped structure
straddling fields F2 and F3.
\label{fig-fovs}
}
\end{figure}

\clearpage

\begin{figure}[p]
\figurenum{2}
\epsscale{0.5}
\plotone{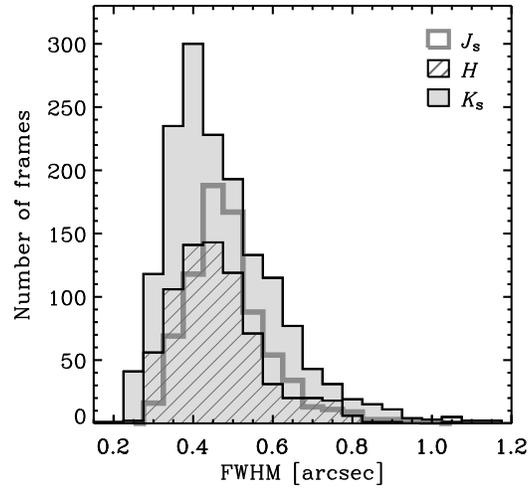}
\vspace{0.0cm}
\caption{
Distributions of the seeing FWHM for the NIR ISAAC observations.
The seeing was measured by fitting Moffat profiles to bright, isolated,
unsaturated stars in the $\rm MS\,1054-03$ individual reduced frames.
Histograms are shown for each band separately:
$J_{\rm s}$ ({\em empty histogram with thick grey line\/}),
$H$ ({\em hatched histogram\/}), and
$K_{\rm s}$ ({\em grey-filled histogram\/}).
\label{fig-seeing}
}
\end{figure}

\clearpage

\begin{figure}[p]
\figurenum{3}
\epsscale{0.98}
\plotone{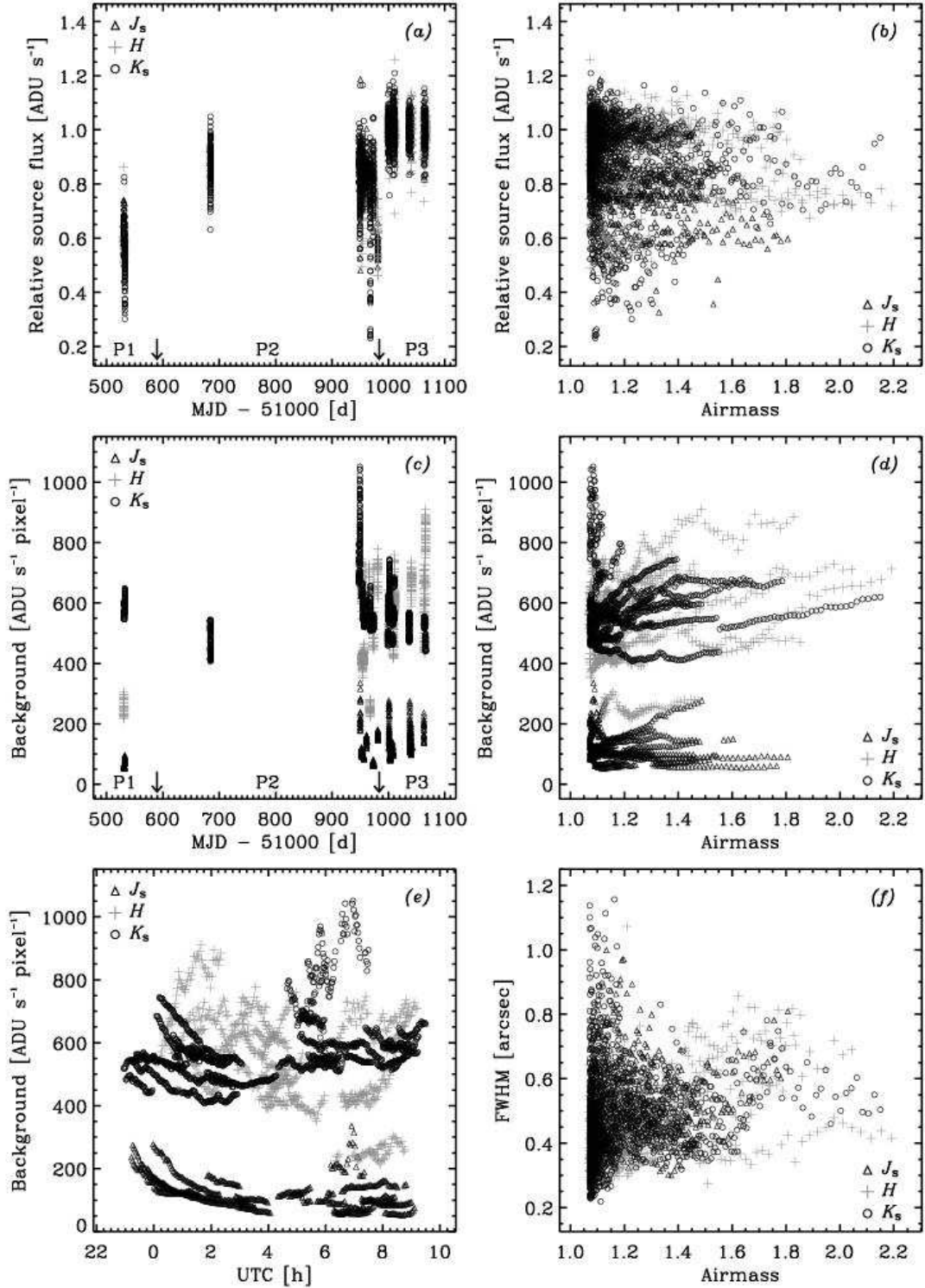}
\vspace{0.0cm}
\caption{
Selected observational parameters for the NIR ISAAC data set.
$J_{\rm s}$-, $H$-, and $K_{\rm s}$-band data are plotted with
{\em black triangles\/}, {\em grey crosses\/}, and {\em black cirlces\/},
respectively.
({\em a\/}) Relative instrumental source count rates derived from bright
stars in the $\rm MS\,1054-03$ field versus modified Julian date (in days).
The arrows at the bottom of the plot indicate the February 2000 and March
2001 VLT Antu primary mirror realuminization, separating the periods P1, P2,
and P3 of data acquisition.
The count rates were integrated in a $6^{\prime\prime}$-diameter
aperture and normalized to the median values over period P3.
({\em b\/}) Relative stars count rates versus airmass.
({\em c\/}) Background levels in the individual raw frames versus modified
Julian date.  The background corresponds to the median count rates in the
raw frames.  Periods P1 to P3 are identified at the bottom of the plot.
({\em d\/}) Background levels versus airmass.
({\em e\/}) Background levels versus Universal Time (in hours).
({\em f\/}) Seeing FWHM of individual frames versus airmass.
In all plots some OBs depart from the general trends, reflecting 
particular conditions during the observations.
\label{fig-obs_param}
}
\end{figure}

\clearpage

\begin{figure}[p]
\figurenum{4}
\epsscale{0.5}
\plotone{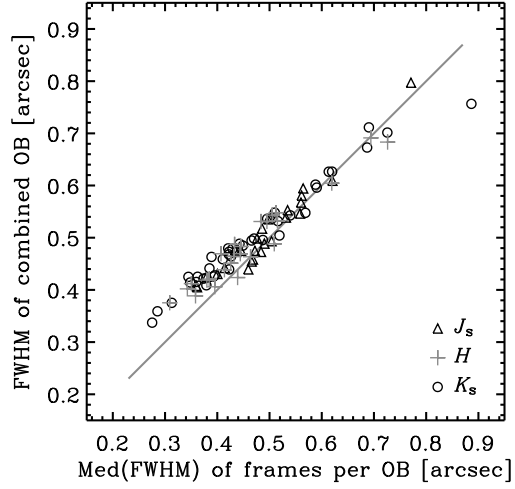}
\vspace{0.0cm}
\caption{
Effects on the PSF of the combination and rectification of the ISAAC OB images.
The FWHM of the PSF measured in the rectified combined OB images is
compared to the median of the FWHMs in the individual non-rectified
frames within each OB.  The angular resolution is preserved to within
$\pm 5\%$ at $\rm FWHM \ge 0\farcs 5$ and degrades progressively
towards lower FWHM by up to 25\% at $\approx 0\farcs 3$, at the
limit of undersampling (see \S~\ref{Sub-red_isaac}).
The FWHM was measured by fitting Moffat profiles to bright, isolated,
unsaturated stars.  Data are shown separately for the three NIR bands:
$J_{\rm s}$: {\em black triangles\/}; $H$: {\em grey crosses\/};
$K_{\rm s}$: {\em black circles\/}.
\label{fig-seeing_OBs}
}
\end{figure}

\clearpage

\begin{figure}[p]
\figurenum{5}
\epsscale{1.0}
\plotone{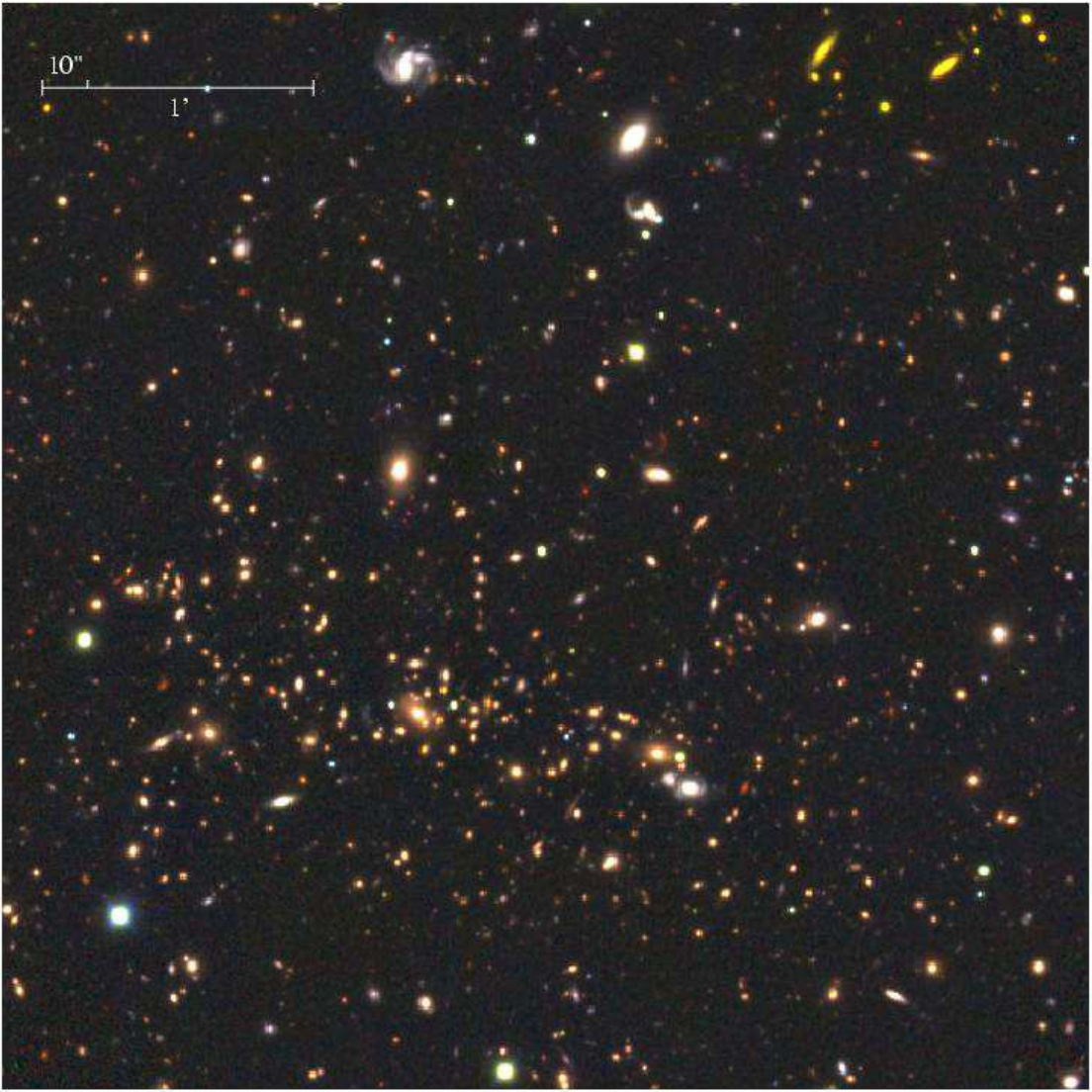}
\vspace{0.0cm}
\caption{
Three-colour composite image of the central
part of the $\rm MS\,1054-03$ field.
The image is constructed from the WFPC2 $I_{814}$ map (coded in blue),
and the ISAAC $J_{\rm s}$ and $K_{\rm s}$ maps (coded in green and red,
respectively).  All three maps are convolved to a common angular resolution
of $\rm FWHM = 0\farcs 69$ (the PSF of the $U$-band map, which has the
poorest resolution among the full data set).  The field of view shown
is $4^{\prime} \times 4^{\prime}$; North is up and East is to the left.
There is a rich variety of colours among the sources, especially for
the fainter ones.  Several of the red objects have spectroscopic or
photometric redshift $z > 2$, and are evolved, dust-obscured, massive
galaxies \citep{FS04}.  Most are extremely faint in the observer's
optical bands and would be missed in optically-selected high-$z$ samples.
The WFPC2 mosaic does not cover the top left and right corners, which
causes the colours of many sources to appear yellow in these regions.
\label{fig-IJK}
}
\end{figure}

\clearpage

\begin{figure}[p]
\figurenum{6}
\epsscale{0.6}
\plotone{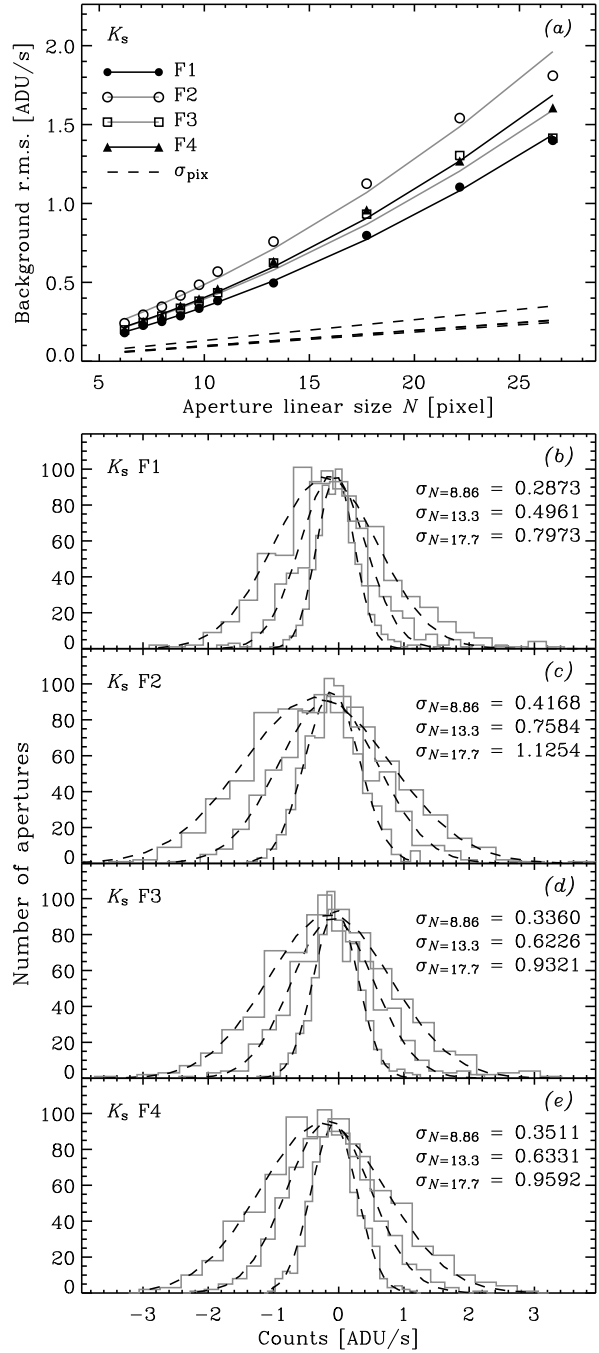}
\vspace{0.0cm}
\caption{
Empirical scaling relation of the background noise rms of the \mscl\ field 
data set as a function of linear size $N = \sqrt{A}$ for apertures of area $A$.
The plots show the relations for the four fields of the ISAAC $K_{\rm s}$-band
mosaic used for the photometric measurements (rectified and resampled to the
WFPC2 scale of $\rm 0\farcs 1~pixel^{-1}$ and smoothed to match the PSF of
the FORS1 $U$-band data with $\rm FWHM = 0\farcs 69$).
The results are qualitatively similar for all other bands.
({\em a\/}) Measured rms variations of the background noise for different
aperture sizes in each of the four ISAAC fields (F1: {\em filled dots\/}; F2:
{\em open dots\/}; F3 : {\em open squares\/}; F4 : {\em filled triangles\/}).
The solid lines indicate the best-fits to the measured relations following
Eq.~\ref{Eq-noise}.  The dashed lines show the linear predictions from the
measured pixel-to-pixel rms if the noise were purely uncorrelated and Gaussian.
({\em b\/}), ({\em c\/}), ({\em d\/}), and ({\em e\/}) Distributions of count
rates measured in random apertures within empty areas of the four ISAAC fields
maps ({\em histograms\/}), which are used to determine the background noise rms
for each aperture size.  The histograms are plotted for three selected aperture
sizes with $N = 8.86$, 13.3, and 17.7 pixel, corresponding to circular
apertures of diameters $d = 1^{\prime\prime}$, $1\farcs 5$, and
$2^{\prime\prime}$.  The background noise rms is derived from the
best-fit Gaussian profile to the distributions ({\em black dashed lines\/}).
\label{fig-noise}
}
\end{figure}

\clearpage

\begin{figure}[p]
\figurenum{7}
\epsscale{0.65}
\plotone{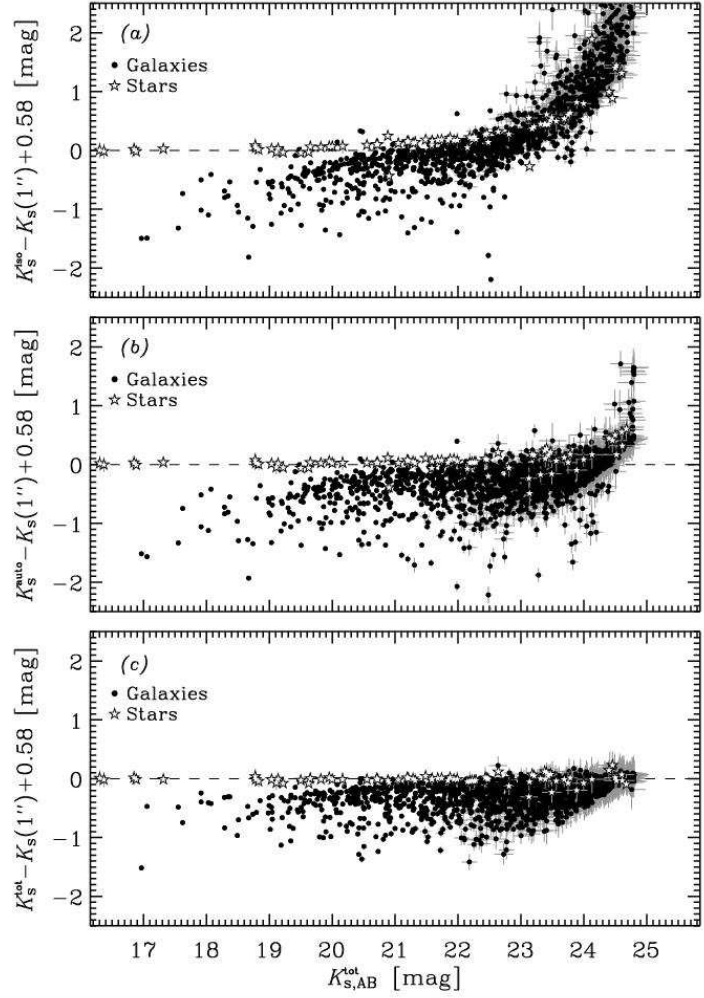}
\vspace{0.0cm}
\caption{
Comparison of various methods to estimate total $K_{\rm s}$-band
magnitudes.
The plots show the difference between the various estimates and the
expected total magnitude for point sources, given by the magnitude
in a 1\farcs 0-diameter circular aperture $K_{\rm s}(1^{\prime\prime})$
scaled by the total aperture correction of 0.58~mag (see \S~\ref{Sub-phot}).
The values of the abcsissa are our adopted total $K_{\rm s}$-band magnitudes.
Stars are distinguished from galaxies ({\em grey stars\/} and
{\em black dots\/}, respectively, see \S~\ref{Sect-redshifts}).
The error bars correspond to $1\sigma$ uncertainties.
For point sources, the best $K^{\rm tot}_{s}$ estimates should
be equal to $K_{\rm s}(1^{\prime\prime}) - 0.58$ over the entire
range in brightness ({\em horizontal black dashed lines\/}).
({\em a\/}) Isophotal magnitudes $K^{\rm iso}_{\rm s}$ strongly
underestimate the total magnitudes $K^{\rm tot}_{\rm s}$ at the faint end
where most sources are point-like and where extended emission lies below
the isophotal surface brightness threshold for detection and photometry.
({\em b\/}) Kron-like autoscaling magnitudes $K^{\rm auto}_{\rm s}$ still
underestimate the $K^{\rm tot}_{\rm s}$ of the fainter objects by up to
0.58~mag, reflecting the emission missed from the PSF wings in the smaller
apertures.
({\em c\/}) The turnover at faint magnitudes is nearly absent
when using the $K^{\rm tot}_{s}$ as measured in this paper.
\label{fig-magtot_comp}
}
\end{figure}

\clearpage

\begin{figure}[p]
\figurenum{8}
\epsscale{0.5}
\plotone{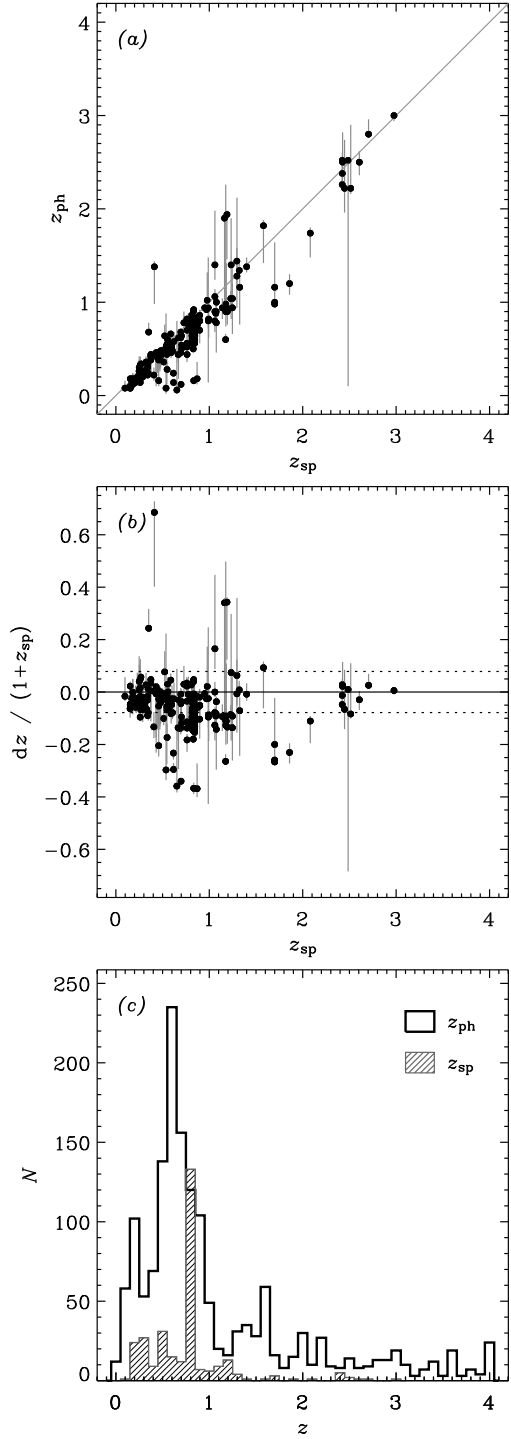}
\vspace{0.0cm}
\caption{
Photometric and spectroscopic redshifts in the $\rm MS\,1054-03$ field.
({\em a\/}) Direct comparison between the photometric redshifts $z_{\rm ph}$
and spectroscopic redshifts $z_{\rm sp}$ for all $K_{\rm s}$-band detected
sources at $K_{\rm s,AB}^{\rm tot} \leq 24.8~{\rm mag}$ for which a
$z_{\rm sp}$ measurement is available.  The vertical error bars indicate
the 68\% confidence intervals on $z_{\rm ph}$ derived from Monte-Carlo
simulations (see \S~\ref{Sect-redshifts}).
In the ideal case, $z_{\rm ph} = z_{\rm sp}$ ({\em grey solid line\/}).
({\em b\/}) Residuals
${\rm d}z / (1+z_{\rm sp}) \equiv (z_{\rm ph}-z_{\rm sp})/(1+z_{\rm sp})$
as a function of spectroscopic redshift.  The accuracy of the photometric
redshifts quantified by the mean $\langle |{\rm d}z| / (1+z_{\rm sp}) \rangle$
is 0.078 ({\em dotted lines\/}) about the ideal value of 0 ({\em solid line\/}).
({\em c\/}) Distributions of photometric redshifts for all $K_{\rm s}$-band
detected sources at $K_{\rm s,AB}^{\rm tot} \leq 24.8~{\rm mag}$
({\em hollow histogram\/}) and of spectroscopic redshifts for the subset
of objects for which they are available ({\rm grey filled histogram}).
In all plots, stars have been excluded.
\label{fig-zph_zsp}
}
\end{figure}

\clearpage

\begin{figure}[p]
\figurenum{9}
\epsscale{0.5}
\plotone{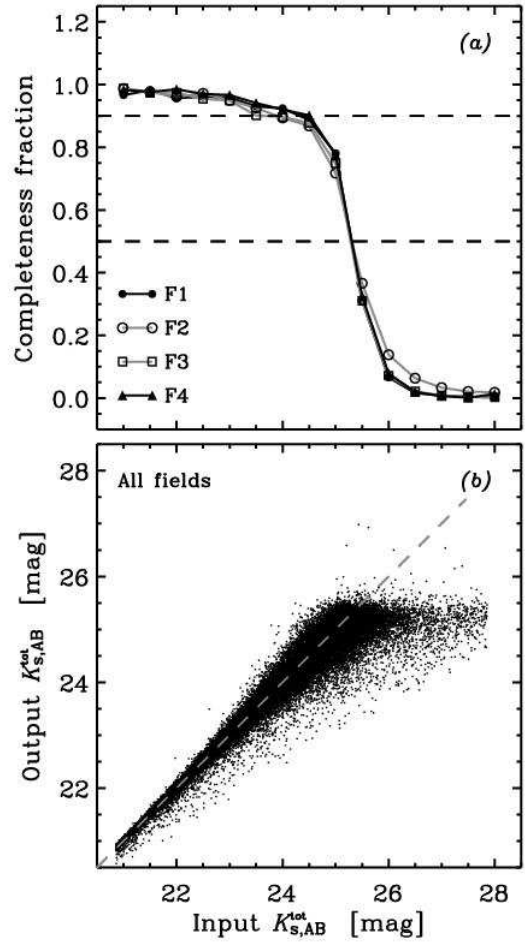}
\vspace{0.0cm}
\caption{
Results of the completeness simulations for point sources in the $K_{\rm s}$
band for the $\rm MS\,1054-03$ field.  For the plots shown here, the analysis
was carried out considering the entire area in each of the ISAAC fields with
effective weight $w_{\rm F} > 0.95$ including also the isophotal areas of
real sources.
({\em a\/}) Completeness curves giving the fraction of recovered
simulated sources as a function of input total $K_{\rm s}$-band magnitude.
The results for each ISAAC field are plotted separately
({\em black line and filled circles\/}: F1;
{\em grey line and open circles\/}: F2;
{\em grey line and open squares\/}: F3;
{\em black line and filled triangles\/}: F4).
The 90\% and 50\% completeness limits are indicated
({\em upper and lower dashed lines}).
({\em b\/}) Observed versus input total $K_{\rm s}$-band magnitude for
all artificial point sources in the simulations in all four ISAAC fields.
The ideal relation where the observed magnitudes equal the input magnitudes
is indicated ({\em grey dashed line\/}).
\label{fig-complete}
}
\end{figure}

\clearpage

\begin{figure}[!t]
\figurenum{10}
\epsscale{0.6}
\plotone{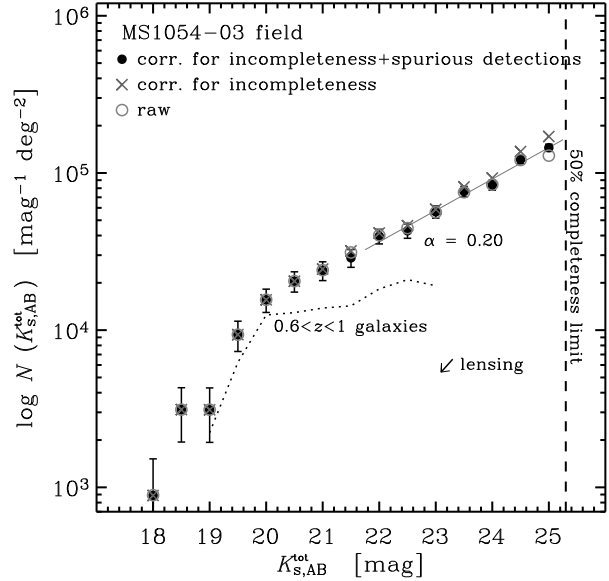}
\vspace{0.0cm}
\caption{
Differential $K_{\rm s}$-band galaxy counts in the $\rm MS\,1054-03$ field.
The counts are given as a function of the total $K_{\rm s}$-band AB magnitude
(with total magnitudes computed as described in \S~\ref{Sub-phot}).
The raw counts ({\em grey open circles\/}) are summed up in bins 0.5~mag wide
(plotted at the central bin values) and over the total area where each ISAAC
field has an effective weight $w_{\rm F}$ exceeding 95\% of its maximum.
The counts were corrected for incompleteness ({\em grey crosses\/}) and for
both incompleteness and spurious detections ({\em black filled circles\/})
using point-source simulations.  Because the four fields have different
noise properties, the corrections were derived and applied for each field
separately, and further account for the variations of effective area as a
function of depth.  The corrections are small except in the faintest bin
near the 50\% completeness or $\approx 3\sigma$ detection limit
({\em dashed vertical line\/}).
Correction for incompleteness only overestimates the counts at the faint
end where the contribution from positive noise peaks becomes important.
The counts at $K_{\rm s,AB}^{\rm tot} \approx 22 - 25~{\rm mag}$
($\rm \approx 20 - 23.5$ in Vega magnitudes) have a logarithmic slope
of $\alpha \approx 0.20$ ({\em grey solid line\/}).
The error bars represent Poisson uncertainties.
Galaxies from the $z = 0.83$ cluster produce the bump seen around
$K_{\rm s,AB}^{\rm tot} \approx 20.5~{\rm mag}$, as indicated by
the number counts of sources around the cluster redshift
({\em dotted line\/}).
Gravitational lensing by the cluster affects little the surface densities:
the effects are small on average over the area surveyed and act along a
direction roughly parallel to the counts curve as illustrated for the
mean magnification by 20\% for $z = 2.5$ background source plane
({\em arrow at bottom left\/}; see \S~\ref{Sub-numcounts}).
\label{fig-numcounts1}
}
\end{figure}


\begin{figure}[!b]
\figurenum{11}
\epsscale{0.6}
\plotone{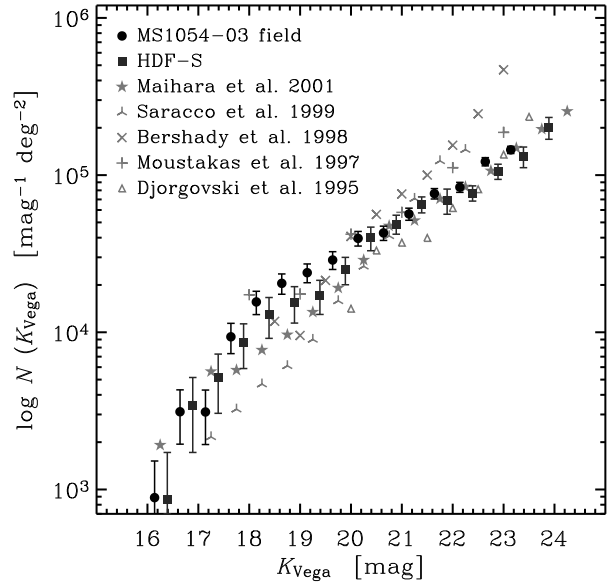}
\vspace{0.0cm}
\caption{
FIRES $K_{\rm s}$-band galaxy counts compared to published counts
in other deep $K$-band fields.  All data are corrected counts, and are
here plotted as a function of magnitudes in the Vega photometric system.
The legend gives the symbols used for different references.
All counts are plotted up to their $\approx 3\sigma$ detection or 50\%
completeness limits, except for the \citet{Sar99} data which extend to
a $\rm S/N \approx 5$.  Error bars are plotted for the FIRES counts and
represent Poisson uncertainties.  The $\approx 5$ times larger area of
the $\rm MS1054-03$ field leads to more robust statistics than for the
deeper but smaller HDF-S up to $K_{\rm Vega} \approx 23~{\rm mag}$, a
depth only about 1~mag brighter than current deep $K$-band surveys.
The cluster bump around $K_{\rm Vega} = 18.5~{\rm mag}$ is conspicuous
for the $\rm MS\,1054-03$ field.
The counts slope at $K_{\rm Vega} > 20~{\rm mag}$ in both FIRES fields
lies generally at the flatter end of results reported in the literature.
The differences between various surveys are probably due in part to
field-to-field variations, and different methods used for the photometric
measurements and for the derivation of the corrections to the raw counts.
\label{fig-numcounts2}
}
\end{figure}

\clearpage

\begin{figure}[p]
\figurenum{12}
\epsscale{0.6}
\plotone{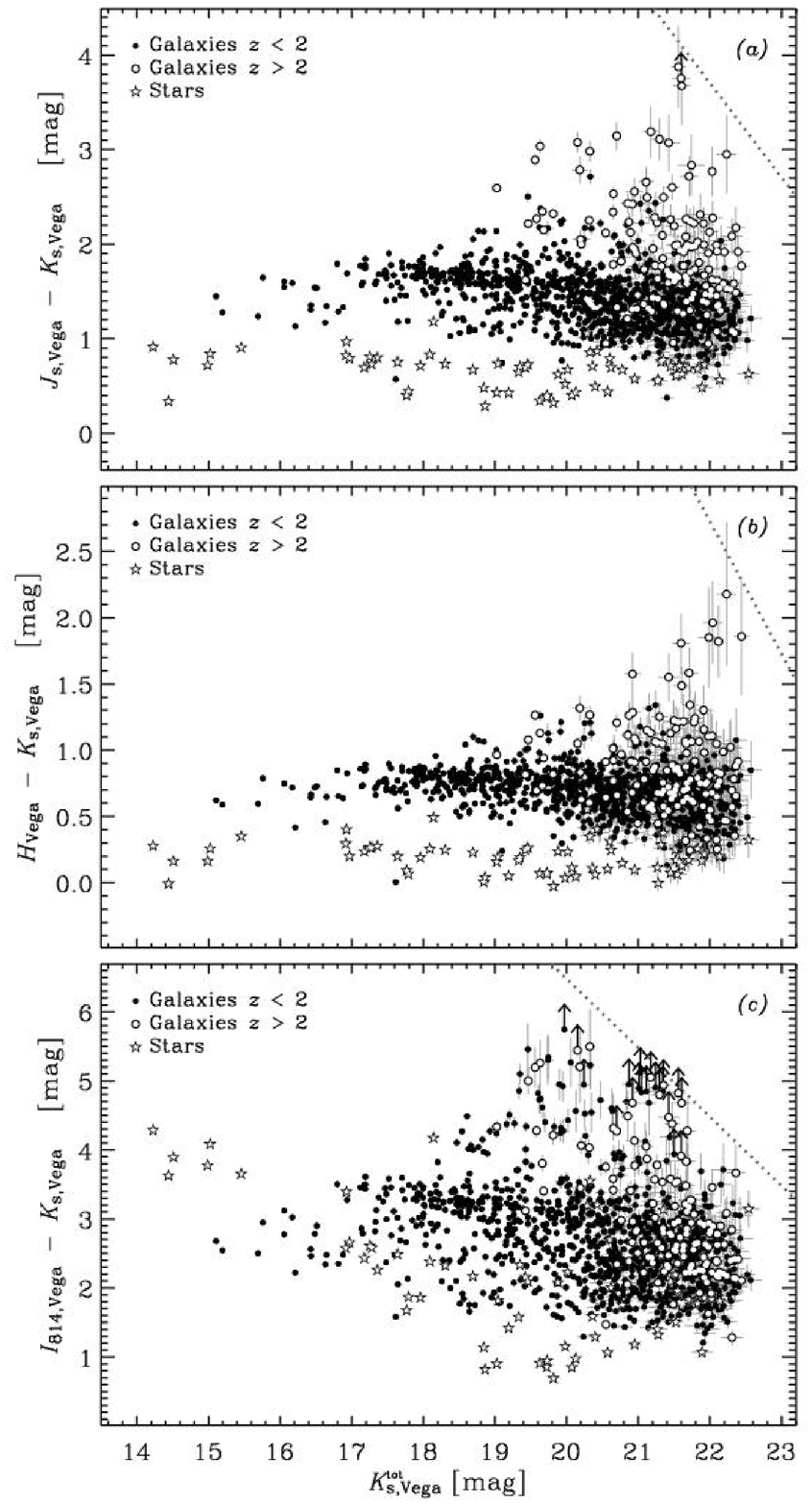}
\vspace{0.0cm}
\caption{
Colour-magnitude diagrams of $K_{\rm s}$-band
selected objects in the $\rm MS\,1054-03$ field.
Magnitudes and colours are on the Vega system.
Only sources with $\rm S/N > 10$ on their $K_{\rm s}$-band colour flux
and a minimum of 30\% of the total exposure time in at least one ISAAC
field for each NIR band and in the $I_{814}$-band map are included.
Galaxies at $z < 2$ and $z > 2$ as well as stars are shown with different
symbols ({\em filled circles\/}, {\em open circles}, and {\em open stars\/},
respectively).
The error bars represent the $1\sigma$ photometric uncertainties,
and sources with $\rm S/N < 2$ in their $J_{\rm s}$-, $H$-, or
$I_{814}$-band fluxes are plotted at their $2\sigma$ confidence level.
The dotted line in each panel indicates the limits corresponding to
the $3\sigma$ background noise level for the point-source aperture
in the PSF-matched images used for the photometry.
({\em a\/}) $J_{\rm s} - K_{\rm s}$ colour
versus total $K_{\rm s}$-band magnitude.
({\em b\/}) $H - K_{\rm s}$ colour
versus total $K_{\rm s}$-band magnitude.
({\em c\/}) $I_{814} - K_{\rm s}$ colour
versus total $K_{\rm s}$-band magnitude.
\label{fig-colmag}
}
\end{figure}

\clearpage

\begin{figure}[p]
\figurenum{13}
\epsscale{0.52}
\plotone{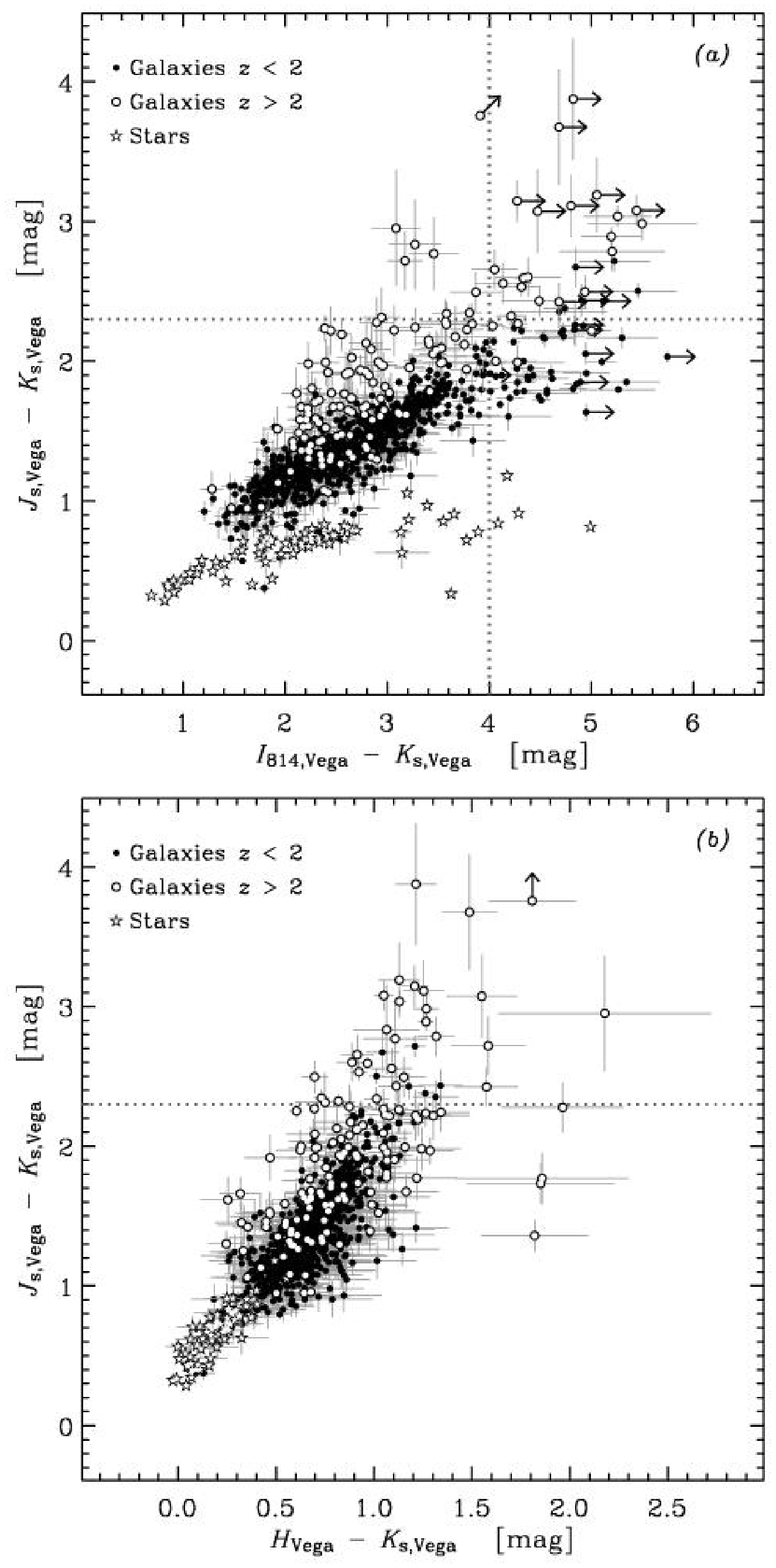}
\vspace{0.0cm}
\caption{
Colour-colour diagrams of $K_{\rm s}$-band
selected objects in the $\rm MS\,1054-03$ field.
The colours are on the Vega system.
Only sources with $\rm S/N > 10$ on their $K_{\rm s}$-band colour flux
and a minimum of 30\% of the total exposure time in at least one ISAAC
field for each NIR band and in the $I_{814}$-band map are included.
Galaxies at $z < 2$ and $z > 2$ as well as stars are shown with different
symbols ({\em filled circles\/}, {\em open circles}, and {\em open stars\/},
respectively).
The error bars represent the $1\sigma$ photometric uncertainties,
and sources with $\rm S/N < 2$ in their $J_{\rm s}$-, $H$-, or
$I_{814}$-band flux are plotted at their $2\sigma$ confidence level.
({\em a\/}) $J_{\rm s} - K_{\rm s}$ versus $I_{814} - K_{\rm s}$ colours.
The criterion $J_{\rm s} - K_{\rm s} > 2.3~{\rm mag}$ defined to identify
Distant Red Galaxies at $z > 2$ \citep[DRGs;][]{Fra03} and the cutoff
$I_{814} - K_{\rm s} > 4.0~{\rm mag}$ to select Extremely Red Objects
(EROs) are indicated ({\em horizontal\/} and {\em vertical dotted lines\/},
respectively).
({\em b\/}) $J_{\rm s} - K_{\rm s}$ versus $H - K_{\rm s}$ colours.
The criterion $J_{\rm s} - K_{\rm s} > 2.3~{\rm mag}$ to select DRGs
at $z > 2$ is again indicated ({\em horizontal dotted line\/}). 
\label{fig-colcol}
}
\end{figure}

\clearpage

\begin{figure}[p]
\figurenum{14}
\epsscale{0.5}
\plotone{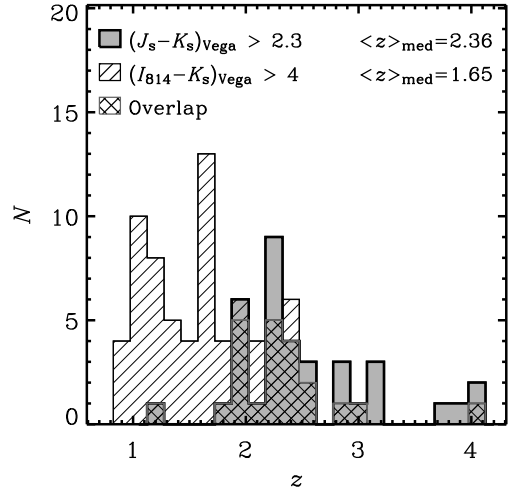}
\vspace{0.0cm}
\caption{
Redshift distributions of red $K_{\rm s}$-band
selected galaxies in the $\rm MS\,1054-03$ field.
Only galaxies with $\rm S/N > 10$ on their $K_{\rm s}$-band colour
flux (implying $K_{\rm s,Vega}^{\rm tot} \leq 22.5~{\rm mag}$) and
a minimum of 30\% of the total exposure time in at least one ISAAC
field for each NIR band and in the $I_{814}$-band map are included.
The spectroscopic redshifts were used wherever available.
The colour cut $J_{\rm s,Vega} - K_{\rm s,Vega} > 2.3~{\rm mag}$
is devised to identify evolved galaxies at $z > 2$ (Distant Red
Galaxies or DRGs; \citealt{Fra03}).  The galaxies that satisfy
this criterion ({\em filled histogram\/}) all lie at $z > 1.8$
except for one interloper at $z = 1.19$, with median redshift
$\langle z \rangle_{\rm med} = 2.36$ (2.42 when excluding the interloper).
The galaxies that have $I_{814,Vega} - K_{\rm s,Vega} > 4$ colours of
Extremely Red Objects (EROs; {\em hatched histogram\/}) typically lie
at lower redshifts, with $\langle z \rangle_{\rm med} = 1.65$.
There is overlap between the two colour-selected samples
({\em cross-hatched histogram\/}) but the
$J_{\rm s,Vega} - K_{\rm s,Vega} > 2.3$ criterion
selects more efficiently $z > 2$ galaxies.
\label{fig-zhist}
}
\end{figure}

\clearpage

\begin{figure}[p]
\figurenum{15}
\epsscale{1.2}
\plotone{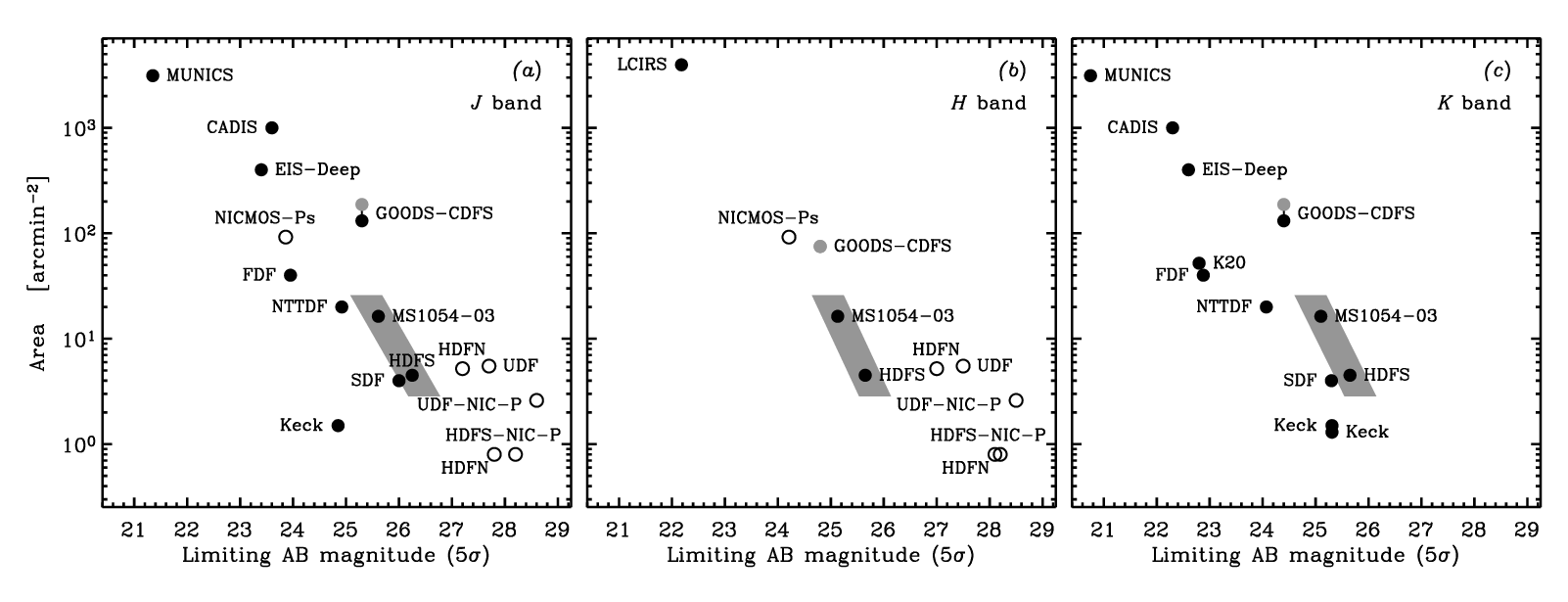}
\vspace{0.0cm}
\caption{
Comparison of published deep NIR imaging surveys in the literature
as a function of depth and area.
({\em a\/}) $J$-band surveys, including HST NICMOS imaging in the
F110W bandpass.
({\em b\/}) $H$-band surveys, including HST NICMOS imaging in the
F160W bandpass.
({\em c\/}) $K$-band surveys, including imaging with $K$, $K_{\rm s}$,
and $K^{\prime}$ filters.
Ground-based and space-based surveys are distinguished by different
symbols ({\em filled circles\/} and {\em open circles\/}, respectively).
In addition to the FIRES HDF-S and \mscl\ fields (highlighted by the
shaded area), the surveys considered include the following:
the NICMOS fields of the HDF-N (HDFN, from \citealt{Tho99} for the
deeper but smaller area, and from \citealt{Dic03} for the wider
and shallower area);
the NICMOS field of the Hubble Ultra Deep Field \citep[UDF,][]{Tho05};
the HDF-S parallel NICMOS field \citep[HDFS-NIC-P,][]{Wil00};
the two UDF parallel NICMOS fields \citep[UDF-NIC-P,][]{Bou05};
the Subaru Deep Field \citep[SDF,][]{Mai01};
several blank fields imaged with the Keck telescopes (three fields in
 $K$ from \citealt{Djo95} and two in $J$ and $K$ from \citealt{Ber98});
the Chandra Deep Field South as part of GOODS \citep[GOODS-CDFS,][]{Gia04};
the ESO New Technology Telescope Deep Field \citep[NTTDF,][]{Sar99};
the FORS Deep Field \citep[FDF,][]{Hei03};
the NICMOS parallel imaging database for Cycle~7 of high Galactic
 latitude fields \citep[NICMOS-Ps,][]{Cor00};
the two fields as basis for the K20 survey \citep[][]{Cim02};
the ESO Imaging Survey Deep fields \citep[EIS-Deep,][]{Arn01};
the Calar Alto Deep Imaging Survey \citep[CADIS,][]{Mei98};
the Las Campanas Infrared Survey \citep[LCIRS,][]{Che02};
the Munich Near-Infrared Cluster Survey \citep[MUNICS,][]{Dro01}.
For GOODS-CDFS, we used the average depth reported by ESO for the
version 1.0 public release of the NIR ISAAC data covering
$\rm 131~arcmin^{2}$ in $J$ and $K_{\rm s}$ (shown in black);
the expected area and additional $H$-band data once the survey is
completed are also indicated (grey filled circle).
As much as possible, the limiting depths refer to total magnitudes,
or magnitudes in fixed apertures significantly larger than the PSF,
for point sources.
The FIRES data of HDF-S currently represents the deepest ground-based
survey in the $J$ and $H$ bands, and the deepest survey in the $K$ band
from either ground or space.
The wider FIRES $\rm MS\,1054-03$ field is the deepest for its size,
filling the regime between the ultra-deep surveys over smaller areas
and significantly larger but yet shallower fields such as GOODS-CDFS.
\label{fig-surveys}
}
\end{figure}

\clearpage

\begin{deluxetable}{cccccc}
\tabletypesize{\small}
\tablecolumns{6}
\tablewidth{360pt}
\tablecaption{Summary of the \mscl\ field observations
              \label{tab-obs}}
\tablehead{
   \colhead{Camera} & 
   \colhead{Filter} & 
   \colhead{Pointings\,\tablenotemark{a}} & 
   \colhead{$N_{\rm exp}$\,\tablenotemark{b}} & 
   \colhead{$t_{\rm int}$\,\tablenotemark{c}} & 
   \colhead{PSF\,\tablenotemark{d}} \\
   \colhead{} & 
   \colhead{} & 
   \colhead{} & 
   \colhead{} & 
   \colhead{(s)} & 
   \colhead{(arcsec)}
}
\startdata
\multicolumn{6}{c}{} \\
\multicolumn{6}{c}{All bands} \\
\cline{1-6} \\
FORS1 & $U$         & 1  &   30 & 16500 & 0.69 \\
FORS1 & $B$         & 1  &   18 &  7200 & 0.57 \\
FORS1 & $V$         & 1  &    6 &  1800 & 0.65 \\
WFPC2 & F606W       & 6  &   12 &  6500 & 0.21 \\
WFPC2 & F814W       & 6  &   12 &  6500 & 0.22 \\
ISAAC & $J_{\rm s}$ & 4  &  776 & 93120 & 0.48 \\
ISAAC & $H$         & 4  &  739 & 87780 & 0.46 \\
ISAAC & $K_{\rm s}$ & 4  & 1588 & 95280 & 0.52 \\
\cline{1-6} \\
\multicolumn{6}{c}{Near-infrared bands} \\
\cline{1-6} \\
ISAAC & $J_{\rm s}$ & F1 &  205 & 24600 & 0.44 \\
      &             & F2 &  211 & 25320 & 0.46 \\
      &             & F3 &  180 & 21600 & 0.45 \\
      &             & F4 &  180 & 21600 & 0.46 \\
ISAAC & $H$         & F1 &  196 & 23520 & 0.46 \\
      &             & F2 &  180 & 20700 & 0.42 \\
      &             & F3 &  184 & 22080 & 0.43 \\
      &             & F4 &  179 & 21480 & 0.42 \\
ISAAC & $K_{\rm s}$ & F1 &  473 & 28380 & 0.46 \\
      &             & F2 &  398 & 23880 & 0.50 \\
      &             & F3 &  358 & 21480 & 0.41 \\
      &             & F4 &  359 & 21540 & 0.39 \\
\enddata
\tablenotetext{a}
{
Number of pointing positions.  
For the ISAAC observations, fields F1, F2, F3, and F4 correspond to the north,
east, south, and west quadrants of the combined mosaic, respectively.  The
properties of the entire mosaic and of each field separately are listed.
}
\tablenotetext{b}
{
Number of individual exposures, or frames.  Frames rejected because of too
poor quality are not included (3\% and 33\% of the total number of exposures
taken with ISAAC and FORS1, respectively).
}
\tablenotetext{c}
{
Total integration time.  Data rejected because of too poor quality are not
included (3\% and 31\% of the total integration time spent with ISAAC and
FORS1, respectively).
}
\tablenotetext{d}
{
FWHM of the best-fitting Moffat profile to the PSF, determined by averaging
the profiles of a sample of bright, isolated, unsaturated stars in the
combined maps.
}
\end{deluxetable}


\begin{deluxetable}{lcc}
\tabletypesize{\small}
\tablecolumns{3}
\tablewidth{280pt}
\tablecaption{Adopted photometric zero points for the \mscl\ data set
              \label{tab-zps}}
\tablehead{
   \colhead{Data set} & 
   \multicolumn{2}{c}{Zero point} \\
   \colhead{} & 
   \colhead{Vega mag} & 
   \colhead{AB mag}
}
\startdata
FORS1 $U$         & 24.127 & 24.828 \\
FORS1 $B$         & 26.746 & 26.634 \\
FORS1 $V$         & 27.131 & 27.145 \\
WFPC2 $V_{606}$   & 22.900 & 23.020 \\
WFPC2 $I_{814}$   & 21.660 & 22.090 \\
ISAAC $J_{\rm s}$ & 24.970 & 25.872 \\
ISAAC $H$         & 24.846 & 26.226 \\
ISAAC $K_{\rm s}$ & 24.309 & 26.169 
\enddata
\tablecomments{
FORS1 and ISAAC measurements were performed in a 6\arcsec -diameter
circular aperture.  WFPC2 values refer to an ``infinite'' aperture,
which is defined as having 1.096 times the flux in an aperture 0\farcs 5
in radius.
The zero points for the FORS1 optical bands were derived applying
$m_{\rm Vega} =
 m_{\rm instr} + {\rm ZP_{Vega}} -
 \kappa \cdot z + \gamma_{\rm C} \cdot {\rm C_{Vega}}$,
where $m_{\rm Vega}$ is the calibrated magnitude of the standard star,
$m_{\rm instr} = -2.5\,\log({\rm counts~rate})$ is the instrumental magnitude,
$z$ is the airmass, and $\rm C$ is the reference colour for the colour term.
The extinction coefficients used were $\kappa(U) = 0.41$, $\kappa(B) = 0.21$,
and $\kappa(V) = 0.16$, in units of $\rm mag\,airmass^{-1}$.  The colour
coefficients were $\gamma_{U-B}(U) = 0.10$, $\gamma_{B-V}(B) = -0.09$, and
$\gamma_{B-V}(V) = 0.05$.  Extinction and colour terms were not applied
for the ISAAC zero points as explained in \S~\ref{Sub-red_isaac}.
The WFPC2 zero points were directly taken from the WFPC2 manual.
}
\end{deluxetable}

\clearpage

\begin{deluxetable}{cccccc}
\tabletypesize{\small}
\tablecolumns{6}
\tablewidth{360pt}
\tablecaption{Background noise in the PSF-matched \mscl\ field images
              \label{tab-noise}}
\tablehead{
   \multicolumn{2}{c}{Data set\,\tablenotemark{a}} & 
   \colhead{$\overline{\sigma}$~\tablenotemark{b}} &
   \colhead{$a$\,\tablenotemark{c}} &
   \colhead{$b$\,\tablenotemark{c}} &
   \colhead{$1\sigma$ background noise limit in 
            $d = 1\farcs 035$\,\tablenotemark{d}} \\
   \colhead{} & 
   \colhead{} & 
   \colhead{(counts/s)} & 
   \colhead{} & 
   \colhead{} & 
   \colhead{(AB mag)}
}
\startdata
FORS1 $U$         &    & $1.13 \times 10^{-3}$ & 1.82 & 0.025 & 29.01 \\
FORS1 $B$         &    & $2.25 \times 10^{-3}$ & 2.31 & 0.093 & 29.60 \\
FORS1 $V$         &    & $9.24 \times 10^{-3}$ & 2.30 & 0.034 & 28.78 \\
WFPC2 $V_{606}$   &    & $1.23 \times 10^{-4}$ & 2.01 & 0.234 & 28.84 \\
WFPC2 $I_{814}$   &    & $1.01 \times 10^{-4}$ & 2.19 & 0.224 & 28.10 \\
ISAAC $J_{\rm s}$ & F1 & 0.00504               & 2.57 & 0.121 & 27.80 \\
                  & F2 & 0.00591               & 2.13 & 0.140 & 27.70 \\
                  & F3 & 0.00531               & 1.96 & 0.143 & 27.87 \\
                  & F4 & 0.00489               & 2.65 & 0.109 & 27.84 \\
ISAAC $H$         & F1 & 0.01054               & 3.29 & 0.101 & 27.20 \\
                  & F2 & 0.01192               & 2.79 & 0.128 & 27.14 \\
                  & F3 & 0.00957               & 3.06 & 0.111 & 27.34 \\
                  & F4 & 0.00939               & 2.71 & 0.136 & 27.40 \\
ISAAC $K_{\rm s}$ & F1 & 0.00926               & 2.45 & 0.128 & 27.45 \\
                  & F2 & 0.01323               & 2.54 & 0.115 & 27.07 \\
                  & F3 & 0.00985               & 2.76 & 0.125 & 27.30 \\
                  & F4 & 0.00975               & 2.73 & 0.143 & 27.28 \\
\enddata
\tablenotetext{a}{
The images are the rectified smoothed maps at $\rm 0\farcs 1~pixel^{-1}$
and common $\rm FWHM = 0\farcs 69$ used for the photometric measurements.
}
\tablenotetext{b}{
Pixel-to-pixel rms as measured directly in empty parts of the images,
within randomly placed circular apertures of 1\arcsec -diameter.
}
\tablenotetext{c}{
Parameters of the best linear fit to the deviations of the effective
background rms variations relative to the uncorrelated Gaussian scaling
as a function of aperture size (\S~\ref{Sub-depths}).
To estimate the photometric uncertainties in a given aperture of linear
size $N = sqrt(A)$ where $A$ is the aperture area, Eq.~\ref{Eq-noise} is
applied with the coefficients listed here and the mean weight within the
aperture.
}
\tablenotetext{d}{
Effective $1\sigma$ background-noise limiting magnitudes of the PSF-matched
images in the point-source aperture appropriate for the convolved images.
The aperture diameter is $d = 1.5 \times {\rm FWHM} = 1\farcs 035$ maximizing
the S/N ratio in the photometry of point sources.
}
\end{deluxetable}


\begin{deluxetable}{ccccc}
\tabletypesize{\small}
\tablecolumns{5}
\tablewidth{360pt}
\tablecaption{Background noise in the non PSF-matched \mscl\ field images
              \label{tab-depth}}
\tablehead{
   \multicolumn{2}{c}{Data set\,\tablenotemark{a}} & 
   \colhead{Point-source aperture diameter\,\tablenotemark{b}} &
   \colhead{$1\sigma$ limiting depth\,\tablenotemark{b}} &
   \colhead{$3\sigma$ total limiting magnitude\,\tablenotemark{c}} \\
   \colhead{} & 
   \colhead{} & 
   \colhead{(arcsec)} &
   \colhead{(AB mag)} &
   \colhead{(AB mag)}
}
\startdata
FORS1 $U$         &    & 1.035 & 29.01 & 27.24 \\
FORS1 $B$         &    & 1.035 & 29.60 & 27.83 \\
FORS1 $V$         &    & 1.035 & 28.78 & 27.01 \\
WFPC2 $V_{606}$   &    & 0.31  & 30.00 & 28.28 \\
WFPC2 $I_{814}$   &    & 0.33  & 29.24 & 27.46 \\
ISAAC $J_{\rm s}$ & F1 & 0.72  & 28.00 & 26.20 \\
                  & F2 & 0.72  & 27.77 & 25.97 \\
                  & F3 & 0.72  & 27.95 & 26.15 \\
                  & F4 & 0.72  & 27.98 & 26.18 \\
ISAAC $H$         & F1 & 0.69  & 27.39 & 25.57 \\
                  & F2 & 0.69  & 27.31 & 25.49 \\
                  & F3 & 0.69  & 27.61 & 25.79 \\
                  & F4 & 0.69  & 27.63 & 25.81 \\
ISAAC $K_{\rm s}$ & F1 & 0.78  & 27.61 & 25.73 \\
                  & F2 & 0.78  & 27.16 & 25.28 \\
                  & F3 & 0.78  & 27.50 & 25.63 \\
                  & F4 & 0.78  & 27.54 & 25.67 \\
\enddata
\tablenotetext{a}{
The images are the rectified maps at $\rm 0\farcs 1~pixel^{-1}$ at their
respective intrinsic angular resolution (listed in Table~\ref{tab-obs}),
except for the FORS1 data.  The $B$- and $V$-band images have a similar
resolution as the $U$-band image used as reference for the PSF-matching
and the differences in the noise analysis from the smoothed and unsmoothed
FORS1 sets can be neglected.
}
\tablenotetext{b}{
Depth of the images as given by the $1\sigma$ background-noise limiting
magnitudes in the point-source apertures appropriate for each image,
with diameter $d = 1.5 \times {\rm FWHM}$.
}
\tablenotetext{c}{
Depth of the images as given by the $3\sigma$ background-noise limiting
total magnitudes, i.e., corrected for the flux missed outside of the aperture
based on the growth curve of the PSFs contructed from bright stars profiles
(see \S~\ref{Sub-phot}).
}
\end{deluxetable}

\clearpage

\begin{deluxetable}{ccccc}
\tabletypesize{\small}
\tablecolumns{5}
\tablewidth{230pt}
\tablecaption{$K_{\rm s}$-band completeness limits for the \mscl\ field
              \label{tab-complete}}
\tablehead{
   \colhead{Field\,\tablenotemark{a}} & 
   \multicolumn{2}{c}{All image area\,\tablenotemark{b}} &
   \multicolumn{2}{c}{Masking real sources\,\tablenotemark{c}} \\
   \colhead{} & 
   \colhead{90\% limit} &
   \colhead{50\% limit} &
   \colhead{90\% limit} &
   \colhead{50\% limit} \\
   \colhead{} & 
   \colhead{(AB mag)} & 
   \colhead{(AB mag)} & 
   \colhead{(AB mag)} & 
   \colhead{(AB mag)}
}
\startdata
F1 & 24.33 & 25.30 & 24.76 & 25.34 \\
F2 & 23.91 & 25.31 & 24.60 & 25.36 \\
F3 & 23.73 & 25.29 & 24.64 & 25.32 \\
F4 & 24.51 & 25.31 & 24.73 & 25.34 \\
\enddata
\tablenotetext{a}{
The $K_{\rm s}$-band image used for the completeness simulations is the
rectified, non PSF-matched image at $\rm 0\farcs 1~pixel^{-1}$ and with
$\rm FWHM = 0\farcs 52$ used for the source detection.
}
\tablenotetext{b}{
Completeness limits derived in simulations where the entire area in each field
with effective weight $w_{\rm F} > 0.95$ is considered, including the regions
within the isophotal areas of real sources.
}
\tablenotetext{c}{
Completeness limits derived in simulations where the area in each field
with effective weight $w_{\rm F} > 0.95$ and outside of the isophotal areas
of real sources is considered.  This reduces the effects of loss and blending
of artificial sources coinciding spatially with real sources.  The impact is
significant for the 90\% limits, which shift towards fainter magnitudes.
}
\end{deluxetable}




\begin{thebibliography}{}

\bibitem[Amico \etal(2002)]{Ami02} Amico, P., Cuby, J.-G., Devillard, N.,
         Yung, Y., \& Lidman, C. 2002, ISAAC Data Reduction Guide, Version 1.5,
         (Garching: ESO)
\bibitem[Appenzeller \etal(1998)]{App98} Appenzeller, I., \etal,
         1998, The Messenger, 94, 1
\bibitem[Arnouts \etal(2001)]{Arn01} Arnouts, S., \etal,
         2001, \aap, 379, 740
\bibitem[Bershady \etal(1998)Bershady, Lowenthal, \& Koo]{Ber98}
         Bershady, M. A., Lowenthal, J. D., \& Koo, D. C.
	 1998, \apj, 505, 50
\bibitem[Bertin \& Arnouts(1996)]{Ber96} Bertin, E., \& Arnouts, S.
         1996, \aaps, 117, 393
\bibitem[Best \etal(2002)]{Bes02} Best, P., van Dokkum, P. G.,
         Franx, M., \& R\"ottgering, H. J. A.
         2002, \mnras, 330, 17
\bibitem[Bouwens \etal(2005)]{Bou05} Bouwens, R. J.,
         Illingworth, G. D., Thompson, R. I., \& Franx, M.
         2005, \apj, 624, L5
\bibitem[Casali \& Hawarden(1992)]{Cas92} Casali, M., \& Hawarden, T.
         1992, JCMT-UKIRT Newsl., No. 4, 33
\bibitem[Chen \etal(2002)]{Che02} Chen, H.-W., \etal,
         2002, \apj, 570, 54
\bibitem[Cimatti \etal(2002)]{Cim02} Cimatti, A., \etal,
         2002, \aap, 392, 395
\bibitem[Corbin \etal(2000)]{Cor00} Corbin, M. R., O'Neil, E.,
         Thompson, R. I., Rieke, M. J., \& Schneider, G.
         2000, \aj, 120, 1209
\bibitem[Daddi \etal(2003)]{Dad03} Daddi, E., \etal,
         2003, \apj, 588, 50
\bibitem[Dickinson \etal(2003)]{Dic03} Dickinson, M., \etal,
         2003, in preparation
\bibitem[Djorgovski \etal(1995)]{Djo95} Djorgovski, S., \etal,
         1995, \apj, 438, L13
\bibitem[Drory \etal(2001)]{Dro01} Drory, N., Feulner, G., Bender, R.,
         Botzler, C. S., Hopp, U., Maraston, C., Mendes de Oliveira, C.,
         \& Snigula, J. 2001, \mnras, 325, 550
\bibitem[F\"orster Schreiber \etal(2004)]{FS04}
         F\"orster Schreiber, N. M., \etal, 2004, \apj, 616, 40
\bibitem[Franx \etal(2000)]{Fra00} Franx, M., \etal, 
         2000, The Messenger, 99, 20
\bibitem[Franx \etal(2003)]{Fra03} Franx, M., \etal, 
         2003, \apj, 587, L79
\bibitem[Giavalisco \etal(2004)]{Gia04} Giavalisco, M., \etal,
         2004, \apj, 600, L93
\bibitem[Hauschildt \etal(1999)]{Hau99} Hauschildt, P. H., Allard, F.,
         Ferguson, J., Baron, E., \& Alexander, D. R.
         1999, \apj, 525, 871
\bibitem[Heidt \etal(2003)]{Hei03} Heidt, J., \etal,
         2003, \aap, 398, 49
\bibitem[Hoekstra \etal(2000)Hoekstra, Franx, \& Kuijken]{Hoe00}
         Hoekstra, H., Franx, M., \& Kuijken, K. 2000, \apj, 532, 88
\bibitem[Jeltema \etal(2001)]{Jel01} Jeltema, T. E., Canizares, C. R., Bautz,
         M. W., Malm, M. R., Donahue, M., \& Garmire, G. P.
	 2001, \apj, 562, 124
\bibitem[Kron(1980)]{Kro80} Kron, R. G.
         1980, \apjs, 43, 305
\bibitem[Kuchinski \etal(2001)]{Kuc01} Kuchinski, L. E., Madore, B. F.,
         Freedman, W. L., \& Trewhella, M.
         2001, \aj, 122, 729
\bibitem[Labb\'e \etal(2003a)]{Lab03a} Labb\'e, I., \etal,
         2003a, \aj, 125, 1107
\bibitem[Labb\'e \etal(2003b)]{Lab03b} Labb\'e, I., \etal,
         2003b, \apj, 591, L95
\bibitem[Labb\'e \etal(2005)]{Lab05} Labb\'e, I., \etal,
         2005, \apj, 624, L81
\bibitem[Landolt(1992)]{Lan92} Landolt, A. U.
         1992, \aj, 104, 340
\bibitem[Madau(1995)]{Mad95} Madau, P. 1995, \apj, 441, 18
\bibitem[Maihara \etal(2001)]{Mai01} Maihara, T., \etal,
         2001, \pasj, 53, 25
\bibitem[McCarthy(2004)]{McC04} McCarthy, P. J.,
         2004, \araa, 42, 477
\bibitem[Meisenheimer(1998)]{Mei98} Meisenheimer, K.
         1998, Astronomische Gesellschaft Meeting Abstracts, 14, 3
\bibitem[Monet \etal(1998)]{Mon98} Monet, D. G., \etal,
         1998, VizieR Online Data Catalog, I/252
\bibitem[Moorwood \etal(1998)]{Moo98} Moorwood, A. F. M., \etal,
         1998, The Messenger, 94, 7
\bibitem[Moustakas \etal(1997)]{Mou97} Moustakas, L. A., Davis, M.,
         Graham, J. R., Silk, J., Peterson, B. A., \& Yoshii, Y.
	 1997, \apj, 475, 445 
\bibitem[Oke(1971)]{Oke71} Oke, J. B. 1971, \apj, 170, 193
\bibitem[Papovich et al.(2001)Papovich, Dickinson, \& Ferguson]{Pap01}
         Papovich, C., Dickinson, M., \& Ferguson, H. C.
	 2001, \apj, 559, 620
\bibitem[Persson \etal(1998)]{Per98} Persson, S. E., Murphy, D. C.,
         Krzeminski, W., Roth, M., \& Rieke, M. J. 1998, \aj, 116, 2475
\bibitem[Rubin \etal(2004)]{Rub04} Rubin, K. H. R., van Dokkum, P. G.,
         Coppi, P., Johnson, O., F\"orster Schreiber, N. M., Franx, M.,
         \& van der Werf, P. 2004, \apj, 613, L5
\bibitem[Rudnick \etal(2001)]{Rud01} Rudnick, G., \etal, 
         2001, \aj, 122, 2205
\bibitem[Rudnick \etal(2003)]{Rud03} Rudnick, G., \etal, 
         2003, \apj, 599, 847 
\bibitem[Saracco \etal(1999)]{Sar99} Saracco, P., D'Odorico, S.,
         Moorwood, A. F. M., Buzzoni, A., Cuby, J.-G., \& Lidman, C.
         1999, \aap, 349, 751
\bibitem[Shapley \etal(2001)]{Sha01} Shapley, A. E., Steidel, C. C.,
         Adelberger, K. L., Dickinson, M., Giavalisco, M., \& Pettini, M.
         2001, \apj, 562, 95
\bibitem[Thompson \etal(1999)]{Tho99} Thompson, R. I.,
         Storrie-Lombardi, L. J., Weymann, R. J., Rieke, M. J.,
	 Schneider, G., Stobie, E., \& Lytle, D.
	 1999, \aj, 117, 17
\bibitem[Thompson \etal(2005)]{Tho05} Thompson, R. I., \etal,
	 2005, \aj, in press (astro-ph/0503504)
\bibitem[Tran \etal(2003)]{Tra03} Tran, K.-V. H., \etal,
	 2003, in preparation
\bibitem[Tran \etal(1999)]{Tra99} Tran, K.-V. H., Kelson, D. D.,
         van Dokkum, P. G., Franx, M., Illingworth, G. D., \& Magee, D.
	 1999, \apj, 522, 39
\bibitem[Trujillo \etal(2003)]{Tru03} Trujillo, I. et al., 
         2003, \apj, 604, 521
\bibitem[Trujillo \etal(2005)]{Tru05} Trujillo, I. et al., 
         2005, \apj, submitted (astro-ph/0504225)
\bibitem[Toft \etal(2005)]{Tof05} Toft, S., van Dokkum, P. G., Franx, M.,
         Thompson, R. I., Illingworth, G. D., Bouwens, R. J., \& Kriek, M.
         2005, \apj, 624, L9
\bibitem[van Dokkum(2001)]{Dok01} van Dokkum, P.G. 2001, \pasp, 113, 1420
\bibitem[van Dokkum \etal(2000)]{Dok00} van Dokkum, P.G., Franx, M.,
         Fabricant, D., Illingworth, G.D., \& Kelson, D.D.
	 2000, \apj, 541, 95
\bibitem[van Dokkum \etal(2005)]{Dok05} van Dokkum, P. G.,
         Kriek, M., Rodgers, B., Franx. M., \& Puxley, P.
         2005, \apj, 622, L13
\bibitem[van Dokkum \etal(2003)]{Dok03} van Dokkum, P. G., \etal 
         2003, \apj, 587, L83
\bibitem[van Dokkum \etal(2004)]{Dok04} van Dokkum, P. G., \etal 
         2004, \apj, 611, 703
\bibitem[Warmels(1991)]{War91} Warmels, R. H., 
         1991, ``The ESO-MIDAS System,''
         in Astronomical Data Analysis Software and Systems I
         (PASP Conf. Series), 25, 115
\bibitem[Williams \etal(2000)]{Wil00} Williams, R. E., \etal,
         2000, \aj, 120, 2735
\bibitem[Zacharias \etal(2004)]{Zac04} Zacharias, N., Urban, S. E.,
         Zacharias, M. I., Wycoff, G. L., Hall, D. M., Germain, M. E.,
         Holdenried, E. R., \& Winter, L.
         2004, \aj, 127, 3043

\end{thebibliography}
\end{document}